\documentclass[11pt]{article}
\usepackage{a4wide}
\usepackage{mathrsfs}
\usepackage{latexsym,bm}
\usepackage{graphicx}
\usepackage{indentfirst}
\usepackage{slashed}
\usepackage{amsmath}
\usepackage{amssymb}
\usepackage[usenames,dvipsnames]{color}
\usepackage{hyperref}
\usepackage{epsfig}
\usepackage[titletoc]{appendix}
\usepackage{multirow}%
\usepackage{rotating}
\usepackage[square,numbers,sort&compress]{natbib}
\usepackage{tikz,tikz-feynman,graphbox,subfigure}   
\usepackage{ulem} 
\usepackage[top=2.9cm,bottom=2.5cm,left=2.8cm,right=3cm]{geometry}
\usepackage{tabularx}

\newcommand{\nn}{\nonumber}

\newcommand{\email}[1]{\footnote{{\em } \texttt{#1}}}

\newcommand{\bma}{\left(\begin{matrix}}
\newcommand{\ema}{\end{matrix}\right)}

\newcommand{\mL}{\mathcal{L}}

\newcommand{\xpt}{{\chi}{\rm PT}}

\begin{document}
\thispagestyle{empty}
\title{
\Large \bf 
Isospin-breaking contribution to the model-independent axion-photon-photon coupling in $U(3)$ chiral theory
}
\author{\small Rui Gao$^a$,\, Jin Hao$^{a}$,\, Chun-Gui Duan$^{a}$\email{duancg@hebtu.edu.cn},\, Zhi-Hui Guo$^{a}$\email{zhguo@hebtu.edu.cn},\, J.~A.~Oller$^{b}$\email{oller@um.es},\,  Hai-Qing Zhou$^{c}$\email{zhouhq@seu.edu.cn} \\[0.5em]
 { \small\it ${}^a$  Department of Physics and Hebei Key Laboratory of Photophysics Research and Application, } \\
 {\small\it Hebei Normal University,  Shijiazhuang 050024, China} \\[0.2em]
{\small {\it ${}^b$ Departamento de F\'{\i}sica. Universidad de Murcia. E-30071 Murcia. Spain}}\\[0.2em]
{ \small\it ${}^c$  School of Physics, Southeast University, Nanjing 211189, China} \\ 
[0.3em]
}
\date{}
  
%

\maketitle
\begin{abstract}
We pursue the calculation of the model-independent component of the axion-photon-photon coupling in the $U(3)$ chiral perturbation theory up to next-to-leading order, with the emphasis on the isospin breaking effect. The mixing of the $\pi^0$-$\eta$-$\eta'$-axion system is revised as well by working out the complete linear isospin-breaking terms. Our calculation shows that the isospin-breaking correction to the axion-photon-photon coupling amounts to more than 15\%, comparing with the result in the isospin limit. 
\end{abstract}

\section{Introduction}

The hypothetical axion, as an ingenious way to address the strong CP problem~\cite{Peccei:1977hh,Weinberg:1977ma,Wilczek:1977pj}, becomes one of the most focused objects not only in particle physics, but also in a broader community, such as astronomy, cosmology, quantum physics, optics, etc~\cite{Kim:2008hd,Graham:2015ouw,Safronova:2017xyt,Irastorza:2018dyq,DiLuzio:2020wdo,Choi:2020rgn,Sikivie:2020zpn}. In the scenario with the axion mass $m_a<1$~MeV, the most promising experimental approach to search for axion relies on the axion-photon-photon interaction~\cite{Sikivie:1983ip,Sikivie:2020zpn}. This latter coupling can be decomposed into the diverse model-dependent part and the unique model-independent part, which correspond to different ultraviolet axion models, such as the Dine-Fischler-Srednicki-Zhitnitsky (DFSZ) and Kim-Shifman-Vainstein-Zakharov (KSVZ)~\cite{Kim:1979if,Shifman:1979if} ones, respectively. In this work, we restrict to the calculation of the model-independent part of the axion-photon-photon coupling. 

The so-called model-independent axion interaction is characterized by the $aG\tilde{G}/f_a$ term, where $G$ and $\tilde{G}$ denote the gluon field strength tensor and its dual, respectively, and $f_a$ stands for the axion decay constant. Chiral perturbation theory ($\xpt$)~\cite{Weinberg:1978kz,Gasser:1983yg,Gasser:1984gg} offers a viable and systematical way to match the $aG\tilde{G}$ term into the hadronic degrees of freedom in the low energy region. Indeed many determinations of the axion-hadron and axion-photon couplings have been given within the $\xpt$ framework, e.g., see some recent works in Refs.~\cite{GrillidiCortona:2015jxo,Ertas:2020xcc,Landini:2019eck,Bigazzi:2019hav,Bauer:2020jbp,Gan:2020aco,DiLuzio:2021vjd,Lu:2020rhp,Vonk:2020zfh,DiLuzio:2022gsc,Wang:2023xny,Gao:2022xqz,Wang:2024tre,Alves:2024dpa,Bai:2024lpq,Cao:2024cym}. 

For the model-independent part of the two-photon axion coupling $g_{a\gamma\gamma}$, its coefficients in units of $\alpha_{em}/(2\pi f_a)$ have been determined to be $1.92\pm0.04$, $2.05\pm0.03$ and $1.63\pm0.01$ in Refs.~\cite{GrillidiCortona:2015jxo,Lu:2020rhp,Gao:2022xqz}, which are obtained in the $SU(2)$, $SU(3)$ and $U(3)$ $\xpt$ up to next-to-leading order (NLO), respectively. The tension between the results of the $U(3)$ and the $SU(2)$, $SU(3)$ frameworks is our concern in this work. Comparing with the approach by rotating away the $aG\tilde{G}/f_a$ via proper quark axial transformation as often used in the $SU(2)$/$SU(3)$ $\xpt$ cases~\cite{Georgi:1986df},  a somewhat different way by explicitly matching the $aG\tilde{G}/f_a$ term into effective operators to incorporate the axion is employed in the $U(3)$ chiral theory in Ref.~\cite{Gao:2022xqz}. This enables the calculation of the mixing formulas of the $\pi^0,\eta,\eta'$ and axion fields order by order in the $U(3)$ $\delta$ counting scheme~\cite{Kaiser:2000gs}, i.e., the simultaneous expansions of momentum, light-quark masses and $1/N_C$: $O(\delta)\sim O(p^2)\sim O(m_q)\sim O(1/N_C)$.  

It is noted that the isospin-breaking (IB) effect is included in the calculation of the $SU(2)$ and $SU(3)$ cases~\cite{GrillidiCortona:2015jxo,Lu:2020rhp}. However, only the leading term in the isospin contributions is kept in Ref.~\cite{Gao:2022xqz}, meaning that for a quantity that does not vanish in the isospin symmetric limit the IB term is simply neglected,  while for a quantity that vanishes in the isospin limit the leading IB correction is then included. The $g_{a\gamma\gamma}$ coupling turns out to correspond to the former category, which is nonzero in the isospin limit with $m_u=m_d$. As a result, the IB correction is not considered for $g_{a\gamma\gamma}$ in the previous $U(3)$ study~\cite{Gao:2022xqz}. The primary aim of this work is to fill this gap by calculating the complete linear IB correction up to NLO in the $\delta$ expansion scheme in $U(3)$ $\xpt$, and to examine its influence on the determination of the axion-photon-photon coupling.

This paper is organized as follows. In Sec.~\ref{sec.mixing} we elaborate the relevant axion-meson $U(3)$ chiral Lagrangian and address the mixing of the $\pi^0$-$\eta$-$\eta'$-axion system. The two-photon couplings for the four particles of $\pi^0$, $\eta$, $\eta'$ and axion are calculated in Sec.~\ref{sec.twophoton}. A short summary and conclusions are presented in Sec.~\ref{sec.concl}.

\section{Axion $U(3)$ $\xpt$ up to next-to-leading order}~\label{sec.mixing}

The way to include the axion in $U(3)$ $\xpt$, relevant chiral Lagrangians, the mixing formulas for the four-particle system of $\pi^0,\eta,\eta'$ and axion, and the calculation of their two-photon couplings, are elaborated in detail in Ref.~\cite{Gao:2022xqz}. A succinct description of the formalism is provided here for the sake of completeness and also for introducing the notations. 

As mentioned in the Introduction, we stick to the model-independent axion interaction with gluons, and the pertinent Lagrangian reads  
\begin{eqnarray}\label{eq.lagag}
\mathcal{L}_a^{G}= \frac{1}{2}\partial_\mu a \partial^\mu a + \frac{a}{f_a} \frac{\alpha_s}{8\pi}G^i_{\mu\nu}\tilde{G}^{i,\mu\nu}  - \frac{1}{2} m_{a,0}^{2}  \, a^2 \,,
\end{eqnarray}
where the sum of the color indices $i$ in the gluon strength tensor $G^i_{\mu\nu}$ and its dual $\tilde{G}^{i,\mu\nu}$ is implicit. The preexisting axion mass $m_{a,0}$ can be considered as a soft breaking of the global $U(1)$ Peccei-Quinn (PQ) symmetry for $m_{a,0}\ll f_a$. For the QCD axion scenario, $m_{a,0}=0$ is imposed and the axion mass is purely generated via the anomalous $aG\tilde{G}$ term in Eq.~\eqref{eq.lagag}. To match this latter term into the $\xpt$ Lagrangian, one popular way is to take a specific axial transformation of the quark fields to eliminate the $aG\tilde{G}$~\cite{Georgi:1986df}, which procedure will introduce two additional pieces: the axion dressed quark mass matrix and the derivative axion interaction with the axial-vector current, e.g., see Ref.~\cite{DiLuzio:2020wdo} for detailed discussions. The two-photon axion coupling in the $SU(2)$~\cite{GrillidiCortona:2015jxo} and $SU(3)$ $\xpt$~\cite{Lu:2020rhp} are calculated by using this procedure. Meanwhile, in the $U(3)$ $\xpt$ which explicitly incorporates the QCD $U_A(1)$ anomaly effect, an alternative way by including the axion together with the singlet $\eta_0$ in its anomalous mass term is employed in Ref.~\cite{Gao:2022xqz}, in order to derive the axion interactions with hadrons and photons. We will follow this latter way to introduce the axion into $U(3)$ $\xpt$ in this work as well. At leading order (LO), the $U(3)$ axion chiral Lagrangian is 
\begin{eqnarray}\label{eq.laglo}
\mL^{\rm LO}= \frac{ F^2}{4}\langle u_\mu u^\mu \rangle+
\frac{F^2}{4}\langle \chi_+ \rangle
+ \frac{F^2}{12}M_0^2 X^2 \,,
\end{eqnarray}
where the singlet $\eta_0$ mass squared $M_0^2$ is proportional to the topological susceptibility, and the axion is incorporated in  
\begin{eqnarray}\label{eq.defx}
X= \log{(\det U)} - i\frac{a}{f_a}\,,  
\end{eqnarray}
and other relevant objects are  
\begin{eqnarray}\label{defbb}
&& U =  u^2 = e^{i\frac{ \sqrt2\Phi}{ F}}\,, \qquad \chi = 2 B M_q \,,\qquad \chi_\pm  = u^\dagger  \chi u^\dagger  \pm  u \chi^\dagger  u \,, \qquad 
 u_\mu = i u^\dagger  \partial_\mu U u^\dagger \,,  
\end{eqnarray} 
with the quark matrix $M_q={\rm diag}(m_u,m_d,m_s)$. To be consistent with the convention of Ref.~\cite{Kaiser:2000gs}, we now use minus sign in front of the axion filed in Eq.~\eqref{eq.defx}, while a plus sign was taken in Ref.~\cite{Gao:2022xqz}. The consequence of this sign change is that the expressions involving one single axion field need to reverse the sign, comparing with the results in Ref.~\cite{Gao:2022xqz}. 
In the $U(3)$ $\xpt$, the singlet $\eta_0$, together with $\pi, K, \eta_8$, are treated as a nonet of pseudo-Nambu-Goldstone bosons (pNGBs)~\cite{Witten:1979vv,Veneziano:1979ec,Coleman:1980mx} and the nonet matrix can be written as 
\begin{equation}\label{phi1}
\Phi \,=\, \left( \begin{array}{ccc}
\frac{1}{\sqrt{2}} \pi^0+\frac{1}{\sqrt{6}}\eta_8+\frac{1}{\sqrt{3}} \eta_0 & \pi^+ & K^+ \\ \pi^- &
\frac{-1}{\sqrt{2}} \pi^0+\frac{1}{\sqrt{6}}\eta_8+\frac{1}{\sqrt{3}} \eta_0   & K^0 \\  K^- & \overline{K}^0 &
\frac{-2}{\sqrt{6}}\eta_8+\frac{1}{\sqrt{3}} \eta_0
\end{array} \right)\,.
\end{equation}
The strong isospin breaking is measured by the factor $\epsilon=B(m_u-m_d)$, with $B$ proportional to the light-quark condensate. In the phenomenological discussion, we will use $\epsilon = m_{K^{+}}^2 - m_{K^0}^2 - (m_{\pi^{+}}^2 - m_{\pi^0}^2)$ to estimate its value, which gives $\epsilon\simeq-5129.2$~MeV$^2$ by taking the physical masses from PDG~\cite{ParticleDataGroup:2022pth}. 

At NLO in the $\delta$-counting scheme, there are four operators~\cite{Gasser:1984gg,Kaiser:2000gs} that are relevant to the mixing of the $\pi^0,\eta,\eta'$ and axion, 
\begin{eqnarray}\label{eq.lagnlo}
\mL^{\rm NLO} =  L_5 \langle u^\mu u_\mu \chi_+ \rangle
+\frac{ L_8}{2} \langle  \chi_+\chi_+ + \chi_-\chi_- \rangle
-\frac{F^2\, \Lambda_1}{12}   D^\mu X D_\mu X  -\frac{F^2\, \Lambda_2}{12} X \langle \chi_- \rangle\,,
\end{eqnarray}
where the low energy constants (LECs) $L_5,L_8,\Lambda_1$ and $\Lambda_2$ are determined in Ref.~\cite{Gao:2022xqz} by fitting the recent lattice results about the $\pi$, $K$, $\eta$ and $\eta'$ parameters as a function of the pion mass.  

In the $U(3)$ $\xpt$ framework, the LO axion-meson and axion-photon interactions, which will be discussed later,  originate from the mixing of the axion and $\pi^0,\eta_8,\eta_0$. As a result, one should first solve the axion-pNGBs mixing problem in the $U(3)$ axion $\xpt$, which can be done order by order in the $\delta$ expansion scheme. In Ref.~\cite{Gao:2022xqz}, the mixing pattern has been worked out for the $\pi^0$-$\eta$-$\eta'$-$a$ system up to NLO by keeping the leading IB contribution, which means that the IB correction is neglected for the quantity which is nonzero in the isospin limit. In order to obtain the full linear IB contribution to the axion-photon-photon coupling, we need to first revise the study for the mixing of $\pi^0,\eta,\eta',a$ discussed in Ref.~\cite{Gao:2022xqz} by completing the calculation of pertinent IB correction terms. 

At LO, there is only mass mixing for the $\pi^0$-$\eta$-$\eta'$-$a$ system, according to the Lagrangian in Eq.~\eqref{eq.laglo}. The diagonalized states at LO, denoted by $\overline{\pi}^0$, $\overline{\eta}$, $\overline{\eta}'$ and $\overline{a}$, can be obtained by taking the field redefinitions as 
\begin{equation}\label{eq.lomat}
\left( \begin{array}{c}
\overline{\pi}^0 \\  \overline{\eta} \\ \overline{\eta}' \\ \overline{a} 
\end{array} \right) \,=\,
 \left( \begin{array}{cccc}
1+v_{11} & -v_{12} & -v_{13} & -v_{14} \\ 
v_{12} & 1+v_{22} & -v_{23} & -v_{24}\\ 
v_{13} & v_{23} & 1+v_{33} & -v_{34} \\ 
v_{41}  & v_{42} & v_{43} & 1+v_{44}
\end{array} \right)\,
\left( \begin{array}{c}
\pi^0 \\  \mathring{\overline{\eta}} \\ \mathring{\overline{\eta}}' \\ a
\end{array} \right) \,, 
\end{equation}
where $\pi^0$ and $a$ stand for the bare states appearing in the chiral Lagrangian, and $ \mathring{\overline{\eta}}$ and $\mathring{\overline{\eta}}'$ correspond to the LO diagonalized states in the isospin limit, which are related to the octet $\eta_8$ and singlet $\eta_0$ through
\begin{eqnarray}\label{eq.loetamixing}
\left(\begin{array}{c} \mathring{\overline{\eta}} \\ \mathring{\overline{\eta}}' \end{array}\right) =
\left(\begin{array}{cc} c_\theta & - s_\theta \\  s_\theta & c_\theta \end{array}\right)  
\left(\begin{array}{c} \eta_8 \\ \eta_0\end{array}\right) \,,
\end{eqnarray}
with $c_\theta=\cos{\theta}$ and $s_\theta=\sin{\theta}$. One can calculate the LO mixing angle $\theta$ and masses for $\mathring{\overline{\eta}}$ and $\mathring{\overline{\eta}}'$ from the Lagrangian~\eqref{eq.laglo}, 
\begin{eqnarray} 
m_{ \mathring{\overline{\eta}}}^2 &=& \frac{M_0^2}{2} + m_{\overline{K}}^2
- \frac{\sqrt{M_0^4 - \frac{4 M_0^2 \Delta^2}{3}+ 4 \Delta^4 }}{2} \,, \label{eq.defmetab2}  \\
m_{\mathring{\overline{\eta}}'}^2 &=& \frac{M_0^2}{2} + m_{\overline{K}}^2
+ \frac{\sqrt{M_0^4 - \frac{4 M_0^2 \Delta^2}{3}+ 4 \Delta^4 }}{2} \,, \label{eq.defmetaPb2}  \\
\sin{\theta} &=& -\left( \sqrt{1 +
\frac{ \big(3M_0^2 - 2\Delta^2 +\sqrt{9M_0^4-12 M_0^2 \Delta^2 +36 \Delta^4 } \big)^2}{32 \Delta^4} } ~\right )^{-1}\,,
\label{eq.deftheta0}
\end{eqnarray}
where $\Delta^2 = m_{\overline{K}}^2 - m_{\overline{\pi}}^2$, and $m_{\overline{\pi}}$, $m_{\overline{K}}$ denote the masses of the pion and kaon in the isospin limit at LO, respectively.

The matrix elements $v_{ij}$ in Eq.~\eqref{eq.lomat} are obtained from the diagonalization procedure of the mass mixing given by the LO Lagrangian~\eqref{eq.laglo}, and the explicit expressions take the form 
\begin{eqnarray}\label{eq.v12}
&v_{12}= -\frac{\epsilon}{\sqrt{3}} \frac{c_\theta-\sqrt{2}s_\theta}{m_{\overline{\pi}}^2-m_{\mathring{\overline{\eta}}}^2}\,, \\
&v_{13}= -\frac{\epsilon}{\sqrt{3}} \frac{\sqrt{2}c_\theta+s_\theta}{m_{\overline{\pi}}^2-m_{\mathring{\overline{\eta}}'}^2}\,, \\
&v_{23}=\frac{\sqrt{2} s_\theta^2+c_\theta s_\theta-\sqrt{2}c_\theta^2}{3(m_{\mathring{\overline{\eta}}'}^2-m_{\mathring{\overline{\eta}}}^2)} \epsilon \,, \\ 
 \label{eq.v24} 
&v_{41}= -\frac{ M_0^2 \epsilon}{6  (m_a^2 - m_{\overline{\pi}}^2)} \frac{F}{f_a}\bigg[-\frac{(\sqrt{2} c_\theta - 
       2 s_\theta) s_\theta}{ m_a^2 - m_{\mathring{\overline{\eta}}}^2} + \frac{c_\theta (2 c_\theta + \sqrt{2} s_\theta)}{ m_a^2 - m_{\mathring{\overline{\eta}}'}^2}\bigg]\,,\\
&v_{14}= -\frac{ M_0^2 \epsilon}{6 (m_a^2 - m_{\overline{\pi}}^2)}\frac{F}{f_a}\bigg[-\frac{(\sqrt{2} c_\theta - 
       2 s_\theta) s_\theta}{ m_{\overline{\pi}}^2 - m_{\mathring{\overline{\eta}}}^2} + \frac{c_\theta (2 c_\theta + \sqrt{2} s_\theta)}{ m_{\overline{\pi}}^2 - m_{\mathring{\overline{\eta}}'}^2}\bigg] \,,\\
&v_{42}= \frac{ M_0^2 s_\theta}{\sqrt{6} (m_a^2 - m_{\mathring{\overline{\eta}}}^2)}\frac{F}{f_a} - \frac{ M_0^2 \epsilon}{ 3 \sqrt{6}  (m_a^2 - m_{\mathring{\overline{\eta}}}^2)}\frac{F}{f_a}\bigg[\frac{c_\theta (-\sqrt{2} c_\theta^2 +  c_\theta s_\theta + \sqrt{2} s_\theta^2)}{ m_a^2 - m_{\mathring{\overline{\eta}}'}^2} - \frac{s_\theta (2  c_\theta^2 + 2\sqrt{2} c_\theta s_\theta +  s_\theta^2)}{m_a^2 - m_{\mathring{\overline{\eta}}}^2}\bigg] \equiv v_{42}^{(0)}+v_{42}^{(1)}\,,\nonumber \\ \\ 
&v_{24}= \frac{ M_0^2 s_\theta}{\sqrt{6}  (m_a^2 - m_{\mathring{\overline{\eta}}}^2)}\frac{F}{f_a} + \frac{ M_0^2 \epsilon}{ 3 \sqrt{6}  (m_a^2 - m_{\mathring{\overline{\eta}}}^2)}\frac{F}{f_a}\bigg[\frac{c_\theta (\sqrt{2} c_\theta^2 -  c_\theta s_\theta -\sqrt{2} s_\theta^2)}{ m_{\mathring{\overline{\eta}}}^2 - m_{\mathring{\overline{\eta}}'}^2} + \frac{s_\theta (2  c_\theta^2 + 2 \sqrt{2} c_\theta s_\theta +  s_\theta^2)}{m_a^2 - m_{\mathring{\overline{\eta}}}^2}\bigg] \equiv v_{24}^{(0)}+v_{24}^{(1)} \,,\nonumber \\ \\  
&v_{43}=- \frac{  M_0^2 c_\theta}{\sqrt{6}  (m_a^2 - m_{\mathring{\overline{\eta}}'}^2)}\frac{F}{f_a} -\frac{ M_0^2 \epsilon}{3 \sqrt{6}  (m_a^2 - m_{\mathring{\overline{\eta}}'}^2)}\frac{F}{f_a} 
\bigg[\frac{c_\theta (c_\theta^2 - 2 \sqrt{2} c_\theta s_\theta + 
2 s_\theta^2)}{m_a^2 - m_{\mathring{\overline{\eta}}'}^2} -\frac{s_\theta (-\sqrt{2} c_\theta^2 + c_\theta s_\theta + \sqrt{2} s_\theta^2)}{m_a^2 - m_{\mathring{\overline{\eta}}}^2}\bigg] \equiv v_{43}^{(0)}+v_{43}^{(1)}\,,\nonumber \\ \\ 
&v_{34}=- \frac{ M_0^2 c_\theta}{\sqrt{6}  (m_a^2 - m_{\mathring{\overline{\eta}}'}^2)} \frac{F}{f_a} + \frac{ M_0^2 \epsilon}{3 \sqrt{6}  (m_a^2 - m_{\mathring{\overline{\eta}}'}^2)}\frac{F}{f_a}
\bigg[-\frac{c_\theta (c_\theta^2 - 2 \sqrt{2} c_\theta s_\theta + 
2 s_\theta^2)}{m_a^2 - m_{\mathring{\overline{\eta}}'}^2} +\frac{s_\theta (-\sqrt{2} c_\theta^2 + c_\theta s_\theta + \sqrt{2} s_\theta^2)}{-m_{\mathring{\overline{\eta}}}^2 + 
m_{\mathring{\overline{\eta}}'}^2}\bigg] \equiv v_{34}^{(0)}+v_{34}^{(1)}\,,\nonumber \\ 
\end{eqnarray}
where $v_{ij}^{(0)}$ denotes the isospin symmetric component, and $v_{ij}^{(1)}$ stands for the linear IB term. For the matrix elements that only contain the IB terms, we do not introduce this decomposition. It is noted that the equalities $v_{42}^{(0)}=v_{24}^{(0)},\, v_{43}^{(0)}=v_{34}^{(0)} $ hold for the isospin symmetric components. For the diagonal $v_{ii}$ elements in Eq.~\eqref{eq.lomat}, they belong to either $O(\epsilon^2)$ or  $O(F^2/f_a^2)$, which therefore will be neglected. 
The masses of the diagonalized fields at LO are found to be 
\begin{equation}
\begin{aligned}
m_{\overline{\eta}}^2=m_{\mathring{\overline{\eta}}}^2+\frac{\epsilon}{3}(\sqrt{2}c_\theta+s_\theta)^2+O(\epsilon^2)\,, 
\end{aligned}
\end{equation}
\begin{equation}
\begin{aligned}
m_{\overline{\eta}'}^2=m_{\mathring{\overline{\eta}}'}^2+\frac{\epsilon}{3}(c_\theta-\sqrt{2} s_\theta)^2+O(\epsilon^2)\,,
\end{aligned}
\end{equation}
\begin{equation}
\begin{aligned}
m_{\overline{a}}^2 =& m_{a,0}^2 + \frac{M_0^2 F^2}{6 f_a^2} \bigg[ 1 + \frac{c_\theta^2 M_0^2 }{m_{a,0}^2 - m_{\mathring{\overline{\eta}}'}^2} +  \frac{s_\theta^2 M_0^2 }{m_{a,0}^2 - m_{\mathring{\overline{\eta}}}^2} \bigg] \\ &+\frac{M_0^4 F^2 \epsilon}{9 f_a^2}  \bigg[\frac{s_\theta^2 (\sqrt{2} c_\theta + s_\theta)^2}{2 (m_{a,0}^2-m_{\mathring{\overline{\eta}}}^2)^2} + \frac{c_\theta^2 (c_\theta - \sqrt{2} s_\theta)^2}{2 (m_{a,0}^2-m_{\mathring{\overline{\eta}}'}^2)^2} + \frac{c_\theta s_\theta (\sqrt{2} c_\theta^2 - c_\theta s_\theta - 
\sqrt{2} s_\theta^2)}{(m_{a,0}^2-m_{\mathring{\overline{\eta}}}^2)( m_{a,0}^2-m_{\mathring{\overline{\eta}}'}^2)}\bigg] + O(\epsilon^2)\,,
\end{aligned}
\end{equation}
and the LO neutral pion mass $m_{\overline{{\pi}}}$ only acquires quadratic IB corrections, which can be disregarded up to the precision of $O(\epsilon)$ implemented in this work.  
By taking $m_{a,0}=0$, one can obtain the axion mass for the conventional QCD axion scenario,  
\begin{equation}
\begin{aligned}
&m_{\overline{a},\rm{QCD~axion}}^2 =\frac{M_0^2 F^2}{6 f_a^2} \bigg[ 1 - \frac{c_\theta^2 M_0^2}{ m_{\mathring{\overline{\eta}}'}^2} -  \frac{s_\theta^2 M_0^2 }{ m_{\mathring{\overline{\eta}}}^2} \bigg] \\ &+\frac{M_0^4 F^2 \epsilon}{9 f_a^2}  \bigg[\frac{s_\theta^2 (\sqrt{2} c_\theta + s_\theta)^2}{2 m_{\mathring{\overline{\eta}}}^4} + \frac{c_\theta^2 (c_\theta - \sqrt{2} s_\theta)^2}{2 m_{\mathring{\overline{\eta}}'}^4} + \frac{c_\theta s_\theta (\sqrt{2} c_\theta^2 - c_\theta s_\theta - 
\sqrt{2} s_\theta^2)}{m_{\mathring{\overline{\eta}}}^2 m_{\mathring{\overline{\eta}}'}^2}\bigg] + O(\epsilon^2)\,. 
\end{aligned}
\end{equation}
It is mentioned that we have corrected a mistake for $v_{14}$ of Ref.~\cite{Gao:2022xqz}, which also affects the axion mass with non-vanishing $m_{a,0}$. However, in the QCD axion case with $m_{a,0}=0$, the expression of the axion mass given in Ref.~\cite{Gao:2022xqz} is not affected. 

As an improvement of the leading IB calculation in Ref.~\cite{Gao:2022xqz}, we now compute the linear IB corrections to the matrix elements $v_{23},v_{24},v_{34},v_{42},v_{43}$ appearing in Eq.~\eqref{eq.lomat} and to the LO masses of $\overline{\eta},\overline{\eta}'$ and axion. In order to give complete linear IB corrections to the axion-photon-photon coupling up to the NLO in the $\delta$ counting scheme, we also need to work out the linear IB expressions for the NLO $\pi^0$-$\eta$-$\eta'$-$a$ mixing. 

The $\overline{\pi}^0,\overline{\eta},\overline{\eta}'$ and $\overline{a}$ fields, which are diagonalized at LO, will get mixed again at NLO, and in Ref.~\cite{Gao:2022xqz} the general bilinear terms involving $\overline{\pi}^0,\overline{\eta},\overline{\eta}'$ and $\overline{a}$ are parameterized as
\begin{equation}\label{eq.lagsenlo}
\begin{aligned}
\mathcal{L}_{\rm NLO}^{\rm bilinear}=&\dfrac{1+\delta^{\eta}_k}{2}\partial_{\mu}\overline\eta\partial^{\mu}\overline\eta+\dfrac{1+\delta^{\eta'}_k}{2}\partial_{\mu}\overline{\eta}'\partial^{\mu}\overline{\eta}'
+\delta_{k}^{\eta\eta'}\partial_{\mu}\overline\eta\partial^{\mu}\overline{\eta}'-\dfrac{m^{2}_{{\overline\eta}}+\delta_{m^{2}_{{\eta}}}}{2}\overline\eta\,\overline\eta\\
&-\dfrac{m^{2}_{{\overline{\eta}'}}+\delta_{m^{2}_{{\eta'}}}}{2}\overline{\eta}'\,\overline{\eta}'-\delta_{m^{2}}^{\eta\eta'}\overline\eta\,\overline{\eta}' +\dfrac{1+\delta^{\pi}_k}{2}\partial_{\mu}\overline{\pi}^{0}\partial^{\mu}\overline{\pi}^{0}+\delta^{\pi\eta}_{k}\partial_{\mu}\overline{\pi}^{0}\partial^{\mu}\overline\eta+\delta^{\pi\eta'}_{k}
\partial_{\mu}\overline{\pi}^{0}\partial^{\mu}\overline{\eta}'\\
&-\dfrac{m_{\overline{\pi}}^{2}+\delta_{m^{2}_{\pi}}}{2}\overline{\pi}^{0}\,\overline{\pi}^{0}-\delta^{\pi\eta}_{m^{2}}\overline{\pi}^{0}\overline{\eta}-\delta^{\pi{\eta}'}_{m^{2}}\overline{\pi}^{0}\overline{\eta}'+\dfrac{1+\delta^{a}_k}{2}\partial_{\mu}\overline{a}\partial^{\mu}\overline{a}+\delta^{a\pi}_{k}\partial_{\mu}\overline{a}\partial^{\mu}\overline{\pi}^{0}+\delta^{a\eta}_{k}\partial_{\mu}\overline{a}
\partial^{\mu}\overline\eta\\
&+\delta^{a{\eta}'}_{k}\partial_{\mu}\overline{a}
\partial^{\mu}\overline{\eta}'-\dfrac{m^{2}_{\overline{a}}+\delta_{m^{2}_{a}}}{2}\overline{a}\,\overline{a}-\delta^{a\pi}_{m^{2}}\overline{a}\,\overline{\pi}^{0}-\delta^{a\eta}_{m^{2}}\overline{a}\,\overline{\eta}-\delta^{a{\eta}'}_{m^{2}}\overline{a}\,\overline{\eta}'\,,
\end{aligned}
\end{equation}
where the NLO corrections are incorporated in the various $\delta_i$ terms, whose explicit expressions after including the complete linear IB contributions are rather lengthy and provided in the Appendix. The diagonalized canonical fields at NLO will be denoted as $\hat{\pi}^0,\hat{\eta},\hat{\eta}'$ and $\hat{a}$, which can be obtained via the following transformation with respect to the LO diagonalized fields 
\begin{align}\label{eq.mixnlo} 
\left( \begin{array}{c}
\hat{\pi}^0 \\   \hat{\eta}  \\  \hat{\eta}' \\ \hat{a} 
\end{array} \right) 
\,=\, &
\left( \begin{array}{cccc}
1   & -y_{12} & -y_{13} & -y_{14} \\ 
y_{12} & 1  & -y_{23} & -y_{24}  \\ 
y_{13} & y_{23} & 1  & -y_{34} \\ 
y_{14} & y_{24} & y_{34} & 1 
\end{array} \right)
\times 
\nonumber \\ & 
\left( \begin{array}{cccc}
1- x_{11} & -x_{12} & -x_{13} & -x_{14} \\ 
-x_{12} & 1- x_{22} & -x_{23} & -x_{24}  \\ 
-x_{13} & -x_{23} & 1-x_{33} & -x_{34} \\ 
-x_{14} & -x_{24} & -x_{34} & 1-x_{44} 
\end{array} \right) \,
\left( \begin{array}{c}
\overline{\pi}^0 \\  \overline{\eta}  \\  \overline{\eta}' \\ \overline{a}
\end{array} \right) \,.
\end{align}
In the above two-step transformation procedure, we first use the matrix with $x_{ij}$ elements to cope with the kinetic mixing, and then use the matrix with $y_{ij}$ elements to handle the remaining mass mixing. The matrix elements $x_{ij}$ and $y_{ij}$ take the forms  
\begin{eqnarray}
&& x_{11}= -\frac{\delta^\pi_k}{2}\,,\quad x_{12}= -\frac{\delta_k^{\pi\eta}}{2}\,,\quad x_{13}= -\frac{\delta_k^{\pi\eta'}}{2}\,,\quad x_{14}= -\frac{\delta_k^{a\pi}}{2}\,, \quad x_{44}= -\frac{\delta^{a}_k}{2}\,, \nonumber\\ && x_{22}= -\frac{\delta^{\eta}_k}{2} \equiv x_{22}^{(0)}+x_{22}^{(1)} \,,  \quad
x_{23}= -\frac{\delta_k^{\eta\eta'}}{2}\equiv x_{23}^{(0)}+x_{23}^{(1)}\,,\quad x_{24}= -\frac{\delta_k^{a\eta}}{2}\equiv x_{24}^{(0)}+x_{24}^{(1)}\,, \nonumber \\ && 
x_{33}= -\frac{\delta^{\eta'}_k}{2}\equiv x_{33}^{(0)}+x_{33}^{(1)}\,,\quad x_{34}= -\frac{\delta_k^{a\eta'}}{2}\equiv x_{34}^{(0)}+x_{34}^{(1)}\,,
\end{eqnarray}
\begin{eqnarray}\label{eq.yij}
&&  y_{12}= \frac{\delta_{m^2}^{\pi\eta}+x_{12}(m_{\overline{\eta}}^2+m_{\overline{\pi}}^2)}{m_{\overline{\eta}}^2-m_{\overline{\pi}}^2}\,, \qquad  y_{13}= \frac{\delta_{m^2}^{\pi\eta'}+x_{13}(m_{\overline{\eta}'}^2+m_{\overline{\pi}}^2)}{m_{\overline{\eta}'}^2-m_{\overline{\pi}}^2}\,, \nonumber\\ &&  y_{14}= \frac{\delta_{m^2}^{a\pi}+x_{14}(m_{a,0}^2+m_{\overline{\pi}}^2)}{m_{a,0}^2-m_{\overline{\pi}}^2}\,,  \qquad 
y_{23}= \frac{\delta_{m^2}^{\eta\eta'}+x_{23}(m_{\overline{\eta}}^2+m_{\overline{\eta}'}^2)}{m_{\overline{\eta}'}^2-m_{\overline{\eta}}^2} \equiv y_{23}^{(0)}+y_{23}^{(1)}\,,\nonumber\\ &&  y_{24}= \frac{\delta_{m^2}^{a\eta}+x_{24}(m_{\overline{\eta}}^2+m_{a,0}^2)}{m_{a,0}^2-m_{\overline{\eta}}^2}\equiv y_{24}^{(0)}+y_{24}^{(1)}\,,\quad y_{34}= \frac{\delta_{m^2}^{a\eta'}+x_{34}(m_{\overline{\eta}'}^2+m_{a,0}^2)}{m_{a,0}^2-m_{\overline{\eta}'}^2}\equiv y_{34}^{(0)}+y_{34}^{(1)}\,, \nonumber \\
\end{eqnarray}
where the superscript $(0)$ denotes the quantity in the isospin limit and the superscript $(1)$ is for the linear IB term. 

The forms for the masses of the diagonalized canonical states at NLO are the same as those given in Ref.~\cite{Gao:2022xqz}, with the obvious replacements of the revised $\delta_i$ including the complete linear IB terms that are given in the Appendix, and we do not repeat them here. By substituting Eqs.~\eqref{eq.lomat} and \eqref{eq.loetamixing} into Eq.~\eqref{eq.mixnlo}, one can get the relations between the physical (diagonalized canonical) states and the bare ones appearing in the Lagrangian, 
\begin{eqnarray}\label{eq.mixnlof08}
 {\tiny  \left( \begin{array}{c}
\hat{\pi}^0 \\   \hat{\eta}  \\  \hat{\eta}' \\ \hat{a} 
\end{array} \right) 
=
\left( \begin{array}{cccc}
1 + z_{11}      & c_\theta(-v_{12}+z_{12}) +s_\theta(-v_{13} + z_{13}) & -s_\theta(-v_{12}+z_{12}) +c_\theta(-v_{13} + z_{13}) & -v_{14}+z_{14} \\ 
v_{12} + z_{21} & c_\theta(1+ z_{22}) +s_\theta ( z_{23} -v_{23}) & -s_\theta(1+ z_{22}) +c_\theta (z_{23}-v_{23})    & -v_{24}+z_{24}  \\ 
v_{13} + z_{31} & c_\theta ( z_{32} + v_{23}) +s_\theta(1+ z_{33})   & -s_\theta (z_{32} + v_{23}) + c_\theta(1+ z_{33})    & -v_{34}+z_{34}  \\ 
v_{41}+z_{41}   & c_\theta(v_{42}+z_{42}) +s_\theta(v_{43}+z_{43})   & -s_\theta(v_{42}+z_{42})+ c_\theta(v_{43}+z_{43}) & 1+ v_{44}+ z_{44} 
\end{array} \right)  
\left( \begin{array}{c}
\pi^0 \\  \eta_8 \\  \eta_0 \\ a
\end{array} \right) }\,, 
\end{eqnarray}
where $z_{ij}$ includes the NLO contributions, 
\begin{eqnarray}
\label{eq.zij}
z_{11}=&& -x_{11}\,, \nonumber \\ 
z_{12}=&& v_{12} x_{11} - x_{12} - y_{12} \,, \nonumber \\
z_{13}=&& v_{13}  x_{11} - x_{13} - y_{13} \,, \nonumber\\
z_{14}=&&   v_{14} x_{11} - x_{14} + v_{42}^{(0)} (x_{12} + y_{12}) +  v_{43}^{(0)} (x_{13} + y_{13}) -  y_{14}\,, \nonumber\\
z_{21}=&&  - x_{12} - v_{12} x_{22}^{(0)} + y_{12} - v_{13} (x_{23}^{(0)} + y_{23}^{(0)}) \,, \nonumber\\
z_{22}=&& - x_{22} - v_{23} (x_{23}^{(0)} + y_{23}^{(0)})\,,\nonumber\\
z_{23}=&& v_{23} x_{22}^{(0)}- x_{23} - y_{23}\,, \nonumber\\ 
z_{24}=&&    v_{24} x_{22}^{(0)} + v_{42}^{(0)} x_{22}^{(1)} -  x_{24} +  v_{34}(x_{23}^{(0)} + y_{23}^{(0)}) +v_{43}^{(0)}(x_{23}^{(1)} + y_{23}^{(1)})-  y_{24}
\,, \nonumber\\ 
z_{31}=&& - x_{13} - v_{12} x_{23}^{(0)} - v_{13} x_{33}^{(0)} + y_{13} + v_{12} y_{23}^{(0)}\,,\nonumber\\ 
z_{32}=&& - x_{23} - v_{23} x_{33}^{(0)} + y_{23} \,, \nonumber\\ 
z_{33}=&& - x_{33} + v_{23} (x_{23}^{(0)} - y_{23}^{(0)})\,,\nonumber\\
z_{34}=&&  v_{24} (x_{23}^{(0)}-y_{23}^{(0)}) +v_{42}^{(0)} (x_{23}^{(1)}-y_{23}^{(1)})+ v_{34} x_{33}^{(0)}+ v_{43}^{(0)} x_{33}^{(1)}  -  x_{34} -  y_{34} \,, \nonumber\\ 
z_{41}=&&   - x_{14} - v_{13} x_{34}^{(0)} + y_{14} + v_{12} ( - x_{24}^{(0)} + y_{24}^{(0)}) + v_{13} y_{34}^{(0)}\,, \nonumber\\ 
z_{42}=&&  - x_{24} + y_{24} + v_{23} (- x_{34}^{(0)} + y_{34}^{(0)}) \,, \nonumber\\ 
z_{43}=&& - x_{34} + v_{23} ( x_{24}^{(0)} - y_{24}^{(0)}) + y_{34} \,,\nonumber\\ 
z_{44}=&& v_{24} (x_{24}^{(0)}-y_{24}^{(0)}) + v_{34} (x_{34}^{(0)}-y_{34}^{(0)})+v_{42}^{(0)} (x_{24}^{(1)}-y_{24}^{(1)}) + v_{43}^{(0)} (x_{34}^{(1)}-y_{34}^{(1)}) - x_{44} \,. \nonumber \\
\end{eqnarray}
Here we have corrected some errors for $z_{ij}$ given in Ref.~\cite{Gao:2022xqz}. 
The low energy constants (LECs) at NLO in Eq.~\eqref{eq.lagnlo} were determined through the fit to various types of lattice QCD data, such as the masses of $\eta$ and $\eta'$, the pion and kaon decay constants, the $\eta$-$\eta'$ mixing parameters, and so on, as a function of $m_\pi$ in Ref.~\cite{Gao:2022xqz}. After including the complete linear IB contributions, we revise the previous fit given in the former reference, which is however confirmed to be barely changed. Therefore in this work we will take the previous fit results in Table 1 of Ref.~\cite{Gao:2022xqz}, which are given below for the sake of completeness,
\begin{eqnarray}
&F=91.05^{+0.42}_{-0.44}~{\rm MeV}\,, \quad L_5= 1.68^{+0.05}_{-0.06}\times 10^{-3}\,, \quad L_8= 0.88^{+0.04}_{-0.04}\times 10^{-3}\,, \nonumber \\  &\Lambda_1= -0.17^{+0.05}_{-0.05}\,, \qquad \Lambda_2= 0.06^{+0.08}_{-0.09} \,. 
\end{eqnarray}

The explicit values of the matrix elements in Eq.~\eqref{eq.mixnlof08} that relate the physical states $\hat{\pi}^0,\hat{\eta},\hat{\eta}',\hat{a}$ and the bare ones $\pi^0,\eta_8,\eta_0,a$ are  
\begin{eqnarray}\label{eq.mixnlof08num}
{\tiny \left( \begin{array}{cccc}
1+ (0.015\pm0.001)     & 0.017 +(-0.007 \pm0.001) & 0.009+ (-0.011 \pm 0.001)  & \frac{-12.8 + (-0.13 \pm 0.02)}{f_a} \\ 
-0.019+(0.005\pm0.001) & 0.94 +(0.21\pm 0.01) & 0.33+(-0.21\pm 0.03) & \frac{-34.3+ (1.7^{+0.8}_{-0.7})}{f_a}  \\ 
-0.003+(-0.001 \pm 0.000) & -0.33 +(-0.18 \pm 0.02) &  0.94+(0.13^{+0.01}_{-0.02})  & \frac{-25.9+(0.2^{+0.4}_{-0.3})}{f_a}  \\ 
\frac{12.1+(0.5\pm 0.1)}{f_a} & \frac{23.8+ (1.0^{+0.2}_{-0.1})}{f_a}  & \frac{35.7+(1.7^{+0.2}_{-0.1})}{f_a} & 1 + \frac{-921.5+(-56.6^{+7.9}_{-9.6})}{f_a^2} 
\end{array} \right)\,,}  \nonumber \\ && 
\end{eqnarray}
where the first entry for each matrix element denotes the LO result, the second one is for the NLO part, and the numbers with $f_a$/$f_a^2$ are given in units of MeV/MeV$^2$. 
To show the convergence of the $\delta$ expansion, we provide the decomposition of the masses up to NLO, 
\begin{eqnarray}~\label{eq.masslo}
m_{\hat\pi}&=& \big[ 134.9 + (0.1 \pm 0.07 ) \big] {\rm MeV}  \,, \nn\\
m_{\hat K}&=& \big[ 492.1 + (5.1^{+3.4}_{-3.3} ) \big] {\rm MeV}  \,, \nn\\
m_{\hat\eta}&=& \big[ 490.4 + (61.1^{+10.0}_{-8.7})  \big] {\rm MeV} \,, \nn\\
m_{\hat{\eta}'} &=& \big[ 954.5+(-28.5^{+11.9}_{-10.9})  \big] {\rm MeV}\,, \nn\\
m_{\hat{a}} &=& \big[5.96 + (0.12\pm 0.02) \big] \mu {\rm eV} \frac{10^{12} {\rm GeV}}{f_a}\,,  
\end{eqnarray}
where the first number for each quantity stands for the LO contribution and the second one is for the NLO part. It is noted that for the results in Eq.~\eqref{eq.masslo} we have taken into account the linear IB corrections, which are however found to be very small, comparing with those in the isospin limit.

\section{Two-photon couplings up to NLO}~\label{sec.twophoton}

In this section, we calculate the two-photon couplings of the axion and pNGBs. They are described by the Wess-Zumino-Witten (WZW) Lagrangian~\cite{Wess:1971yu,Witten:1983tw}, and at LO it only consists of one operator, 
\begin{eqnarray}\label{eq.lagwzwlo}
\mathcal{L}_{WZW}^{\rm LO}= -\frac{3\sqrt{2}}{8\pi^2 F}\varepsilon_{\mu\nu\rho\sigma}\partial^\mu A^\nu \partial^\rho A^\sigma \langle Q^2 \Phi \rangle  \,,
\end{eqnarray}
with $Q={\rm Diag}(\frac{2e}{3},-\frac{e}{3},-\frac{e}{3})$ the electric charge matrix of the light quarks $u,d,s$. At NLO in the $\delta$ counting scheme, two additional operators appear~\cite{Moussallam:1994xp,Bijnens:2001bb} and their relevant forms to our study read 
\begin{align}\label{eq.lagwzwnloexpd}
\mathcal{L}_{WZW}^{\rm NLO}=&  t_1 \frac{32\sqrt{2} B }{F}\varepsilon_{\mu\nu\rho\sigma}\partial^\mu A^\nu \partial^\rho A^\sigma \langle \big( M_q \Phi + \Phi M_q  \big) Q^2 \rangle \nonumber \\ &+
   16  k_3 \varepsilon_{\mu\nu\rho\sigma}\partial^\mu A^\nu \partial^\rho A^\sigma \langle Q^2 \rangle \bigg( \frac{\sqrt{2}}{F} \langle \Phi \rangle - \frac{a}{f_a} \bigg)\,.
\end{align}

With the forms of the $\pi^0$-$\eta$-$\eta'$-axion mixing at NLO obtained in the previous section, we are ready to derive the axion-photon-photon couplings from the WZW chiral Lagrangians in Eqs.~\eqref{eq.lagwzwlo} and \eqref{eq.lagwzwnloexpd}. To be more specific, one can substitute the inverse of Eq.~\eqref{eq.mixnlof08} into the WZW Lagrangians~\eqref{eq.lagwzwlo} and \eqref{eq.lagwzwnloexpd}, and then expand the chiral operators in terms of the physical states up to NLO. From this procedure, the LO WZW Lagrangian in Eq.~\eqref{eq.lagwzwlo} gives 
\begin{equation}\label{eq.lagwzwlophys}
\begin{aligned}
&\mathcal{L}^{\rm LO}_{\rm WZW}=\frac{e^2}{24 F \pi^2}\epsilon_{\mu\nu \rho \sigma}\partial^\mu A^\nu \partial^\rho A^\sigma \Bigg\{-\bigg[3 v_{41} - 2 \sqrt{6} s_\theta v_{42} +  \sqrt{3} s_\theta v_{43} + 3 x_{14} - 2 \sqrt{6} s_\theta x_{24}\\ & + 3 v_{12} x_{24}^{(0)} - \sqrt{3} s_\theta v_{23} x_{24}^{(0)} + \sqrt{3} s_\theta x_{34} + 3 v_{13} x_{34}^{(0)} - 2 \sqrt{6} s_\theta v_{23} x_{34}^{(0)} + 3 y_{14} - 2 \sqrt{6} s_\theta y_{24} + 3 v_{12} y_{24}^{(0)}\\ & - \sqrt{3} s_\theta v_{23} y_{24}^{(0)} + \sqrt{3} s_\theta y_{34}+ 3 v_{13} y_{34}^{(0)} - 2 \sqrt{6} s_\theta v_{23} y_{34}^{(0)} + \sqrt{3} c_\theta (v_{42} + 2 \sqrt{2} v_{43} \\ & + x_{24} - 2 \sqrt{2} v_{23} x_{24}^{(0)} + 2 \sqrt{2} x_{34} + v_{23} x_{34}^{(0)} + y_{24} - 2 \sqrt{2} v_{23} y_{24}^{(0)} + 2 \sqrt{2} y_{34} + v_{23} y_{34}^{(0)})\bigg] \hat{a} \\ & + \bigg[-3 (1 + x_{11}) + \sqrt{3} s_\theta \big[v_{13} + v_{13} x_{11} - 2 \sqrt{2} v_{12} (1 + x_{11}) + 2 \sqrt{2} x_{12} - x_{13} - 2 \sqrt{2} y_{12} + y_{13}\big] \\ &+ \sqrt{3} c_\theta (v_{12} (1 + x_{11}) + 2 \sqrt{2} v_{13} (1 + x_{11}) - x_{12} - 2 \sqrt{2} x_{13} + y_{12} + 2 \sqrt{2} y_{13})\bigg]\hat{\pi}^0  \\ &+ \bigg[-3 (v_{12} + x_{12} + v_{12} x_{22}^{(0)} + v_{13} x_{23}^{(0)} + y_{12} - v_{13} y_{23}^{(0)}) + \sqrt{3} c_\theta \big[-1 - x_{22} - 2 \sqrt{2} x_{23} + 2 \sqrt{2} y_{23} \\ &+ v_{23} (2 \sqrt{2} + 2 \sqrt{2} x_{22}^{(0)} - x_{23}^{(0)} + y_{23}^{(0)})\big]\\ & + \sqrt{3} s_\theta \big[ 2 \sqrt{2} + 2 \sqrt{2} x_{22} - x_{23} + y_{23} + v_{23} (1 + x_{22}^{(0)} + 2 \sqrt{2} x_{23}^{(0)} - 2 \sqrt{2}y_{23}^{(0)})\big]\bigg] \hat{\eta} \\ & -  \bigg[3 (v_{13} + x_{13} + v_{12} x_{23}^{(0)} + v_{13} x_{33}^{(0)} + y_{13} + v_{12} y_{23}^{(0)}) - \sqrt{3} s_\theta \big[-1 + 2 \sqrt{2} x_{23} - x_{33} + 2 \sqrt{2} y_{23} \\ &+ v_{23} (2 \sqrt{2} + x_{23}^{(0)} + 2 \sqrt{2} x_{33}^{(0)} + y_{23}^{(0)})\big]\\ & + \sqrt{3} c_\theta \big[2 \sqrt{2} + x_{23} + 2 \sqrt{2} x_{33} + y_{23} + v_{23} (1 - 2 \sqrt{2} x_{23}^{(0)} + x_{33}^{(0)} - 2 \sqrt{2} y_{23}^{(0)})\big]\bigg] \hat{\eta}'\Bigg\}\,, 
\end{aligned}
\end{equation}
which now includes partial NLO contributions through the mixing in Eq.~\eqref{eq.mixnlof08},  
and the NLO WZW Lagrangian in Eq.~\eqref{eq.lagwzwnloexpd} leads to 
\begin{equation}\label{eq.lagwzwnlophys}
\begin{aligned}
&\mathcal{L}^{\rm NLO}_{\rm WZW}=\frac{e^2}{27 F } \epsilon_{\mu\nu \rho \sigma}\partial^\mu A^\nu \partial^\rho A^\sigma \Bigg\{32 \hat{a}\big\{- 9  k_3 \frac{F}{f_a}+ \bigg[9 \sqrt{6} k_3 (-s_\theta v_{42} + c_\theta v_{43}) \\ &+ m_\pi^2 t_1 \big[9 v_{41} + \sqrt{3} (7 c_\theta v_{42} - 4 \sqrt{2} s_\theta v_{42} + 4 \sqrt{2} c_\theta v_{43} + 7 s_\theta v_{43})\big] + \sqrt{3} t_1 \big[-2 m_K^2 s_\theta (\sqrt{2} v_{42} + 2 v_{43})\\ & + c_\theta m_K^2 (-4 v_{42} + 2 \sqrt{2} v_{43})  +s_\theta (-4 \sqrt{2} v_{42}^{(0)} + v_{43}^{(0)}) \epsilon + c_\theta (v_{42}^{(0)} + 4 \sqrt{2} v_{43}^{(0)}) \epsilon\big]\bigg] \big\}\\ & +  32  \Big\{9 \sqrt{6} k_3 (s_\theta v_{12} - c_\theta v_{13}) + m_\pi^2 t_1 (9 - 7 \sqrt{3} c_\theta v_{12} + 4 \sqrt{6} s_\theta v_{12} - 4 \sqrt{6} c_\theta v_{13} - 7 \sqrt{3} s_\theta v_{13}) \\ &+ t_1 \big[2 \sqrt{3} m_K^2 s_\theta (\sqrt{2} v_{12} + 2 v_{13}) + 2 \sqrt{3} c_\theta m_K^2 (2 v_{12} - \sqrt{2} v_{13}) + 
15 \epsilon \big]\Big\} \hat{\pi}^0 \\ &- 32\Big\{9 \sqrt{6} k_3 (s_\theta + c_\theta v_{23}) + t_1 \big[4 \sqrt{6} m_\pi^2 s_\theta - 9 m_\pi^2 v_{12} + 2 \sqrt{3} m_K^2 s_\theta (\sqrt{2} - 2 v_{23}) + 7 \sqrt{3} m_\pi^2 s_\theta v_{23}\\ & + \sqrt{3} c_\theta (2 m_K^2 (2
+ \sqrt{2} v_{23}) + m_\pi^2 (-7 + 4 \sqrt{2} v_{23}) - \epsilon) + 
4 \sqrt{6} s_\theta \epsilon\big]\Big\} \hat{\eta} \\ & + 32  \Big\{9 \sqrt{6} c_\theta k_3 + 7 \sqrt{3} m_\pi^2 s_\theta t_1 + 9 m_\pi^2 t_1 v_{13} - 9 \sqrt{6} k_3 s_\theta v_{23} - 
4 \sqrt{6} m_\pi^2 s_\theta t_1 v_{23} + \sqrt{3} s_\theta t_1 \epsilon\\ & - 2 \sqrt{3} m_K^2 s_\theta t_1 (2 + \sqrt{2} v_{23}) + \sqrt{3} c_\theta t_1 \big[2 m_K^2 (\sqrt{2} - 2 v_{23}) + m_\pi^2 (4 \sqrt{2} + 7 v_{23}) + 4 \sqrt{2} \epsilon\big]\Big\} \hat{\eta}'\Bigg\}\,, 
\end{aligned}
\end{equation}
where only the NLO pieces are retained. 

By taking the WZW Lagrangians~\eqref{eq.lagwzwlophys} and \eqref{eq.lagwzwnlophys} written in terms of the physical states, we can calculate  the two-photon decay amplitudes of $\phi\to\gamma(k_1)\gamma(k_2)$, 
\begin{eqnarray}\label{eq.defTphigg}
 T_{\phi\to \gamma\gamma} = e^2\varepsilon_{\mu\nu\rho\sigma} k_1^\mu \epsilon_1^\nu k_2^\rho \epsilon_2^{\sigma} F_{\phi\gamma\gamma} \,, 
\end{eqnarray}
where $\phi$ stands for the physical states of $\hat{\pi}^0,\hat{\eta},\hat{\eta}',\hat{a}$ that are diagonalized up to NLO, and $\epsilon_{1,2}$ correspond to the polarization vectors of the photons, and the two-photon couplings include both LO and NLO contributions, i.e., $F_{\phi\gamma\gamma}=F_{\phi\gamma\gamma}^{\rm LO}+F_{\phi\gamma\gamma}^{\rm NLO}$. In the following, we will simply use the notations of $\pi^0,\eta,\eta',a$ for the physical states $\hat{\pi}^0,\hat{\eta},\hat{\eta}',\hat{a}$. 
The two-photon couplings for $\pi^0,\eta,\eta',a$ at LO and NLO take the form 
\begin{equation}
\begin{aligned}
F_{\pi^0\gamma\gamma}^{\rm LO}&=-\frac{1}{12 F \pi^2}\big[-3+\sqrt{3}s_\theta(v_{13}-2\sqrt{2}v_{12})+\sqrt{3}c_\theta(v_{12}+2\sqrt{2}v_{13})\big]\,, \\
F_{\pi^0\gamma\gamma}^{\rm NLO}&= -\frac{1}{12 F \pi^2}\Bigg\{\bigg[-3  x_{11}+ \sqrt{3} s_\theta \big[ v_{13} x_{11} - 2 \sqrt{2} v_{12}  x_{11} + 2 \sqrt{2} x_{12} - x_{13}  -
2 \sqrt{2} y_{12} + y_{13}\big] \\ &+ \sqrt{3} c_\theta \big[v_{12}  x_{11} +
2 \sqrt{2} v_{13}  x_{11} - x_{12} - 2 \sqrt{2} x_{13} + y_{12} +
2 \sqrt{2} y_{13}\big]\bigg]\Bigg\} \\ &- \frac{64}{27 F}\Bigg\{ 9 m_\pi^2 t_1 +
9 \sqrt{6} k_3 s_\theta v_{12} + 4 \sqrt{6} m_\pi^2 s_\theta t_1 v_{12} - 9 \sqrt{6} c_\theta k_3 v_{13} - 7 \sqrt{3} m_\pi^2 s_\theta t_1 v_{13} + 15 t_1 \epsilon \\
& + 2 \sqrt{3} m_K^2 s_\theta t_1 (\sqrt{2} v_{12} + 2 v_{13}) + \sqrt{3} c_\theta t_1 \bigg[m_\pi^2 (-7 v_{12} - 4 \sqrt{2} v_{13}) - 2 m_K^2 (-2 v_{12} + \sqrt{2} v_{13})\bigg] \Bigg\}\,,
\end{aligned}
\end{equation}
\begin{equation}
\begin{aligned}
F_{\eta\gamma\gamma}^{\rm LO}&=-\frac{1}{12 F \pi^2}\big[-3v_{12}+\sqrt{3}c_{\theta}(-1+2\sqrt{2}v_{23})+\sqrt{3}s_{\theta}(2\sqrt{2}+v_{23})\big]\,,  \\ 
F_{\eta\gamma\gamma}^{\rm NLO}&=-\frac{1}{12 F \pi^2} \Bigg\{-3 ( x_{12} + v_{12} x_{22}^{(0)} + v_{13} x_{23}^{(0)} + y_{12} - v_{13} y_{23}^{(0)}) \\ & + \sqrt{3} c_\theta \bigg[-x_{22} - 2 \sqrt{2} x_{23} + 2 \sqrt{2} y_{23} +
v_{23} ( 2 \sqrt{2} x_{22}^{(0)} - x_{23}^{(0)} + y_{23}^{(0)})\bigg] \\ &+ \sqrt{3} s_\theta \bigg[ 2 \sqrt{2} x_{22} - x_{23} +
y_{23}  + v_{23} ( x_{22}^{(0)} + 2 \sqrt{2} x_{23}^{(0)} - 2 \sqrt{2} y_{23}^{(0)})\bigg]\Bigg\} \\ &- \frac{64}{27 F}  \Bigg\{-9 \sqrt{6} k_3 s_\theta - 4 \sqrt{6} m_\pi^2 s_\theta t_1 + 9 m_\pi^2 t_1 v_{12} - 2 \sqrt{3} m_K^2 s_\theta t_1 (\sqrt{2} - 2 v_{23}) -
9 \sqrt{6} c_\theta k_3 v_{23} \\ &- 7 \sqrt{3} m_\pi^2 s_\theta t_1 v_{23} -
4 \sqrt{6} s_\theta t_1 \epsilon + \sqrt{3} c_\theta t_1 \bigg[m_\pi^2 (7 - 4 \sqrt{2} v_{23}) - 2 m_K^2 (2 + \sqrt{2} v_{23}) + \epsilon\bigg]\Bigg\} \,,
\end{aligned}
\end{equation}
\begin{equation}
\begin{aligned}
F_{\eta'\gamma\gamma}^{\rm LO}&=\frac{1}{12 F \pi^2}\big[3v_{13}-\sqrt{3}s_{\theta}(-1+2\sqrt{2}v_{23})+\sqrt{3}c_{\theta}(2\sqrt{2}+v_{23})\big] \,, \\ 
F_{\eta'\gamma\gamma}^{\rm NLO}&= \frac{1}{12 F \pi^2} \Bigg\{3 ( x_{13} + v_{12} x_{23}^{(0)} + v_{13} x_{33}^{(0)} + y_{13} + v_{12} y_{23}^{(0)}) \\ &- \sqrt{3} s_\theta \bigg[ 2 \sqrt{2} x_{23} - x_{33} + 2 \sqrt{2} y_{23}  + v_{23} ( x_{23}^{(0)} + 2 \sqrt{2} x_{33}^{(0)} + y_{23}^{(0)})\bigg] \\ &+ \sqrt{3} c_\theta \bigg[ x_{23} + 2 \sqrt{2} x_{33} + y_{23}  + v_{23} ( - 2 \sqrt{2} x_{23}^{(0)} + x_{33}^{(0)} - 2 \sqrt{2} y_{23}^{(0)})\bigg]\Bigg\}\\ &- \frac{64}{27 F} \Bigg\{9 \sqrt{6} c_\theta k_3 + 7 \sqrt{3} m_\pi^2 s_\theta t_1 + 9 m_\pi^2 t_1 v_{13} - 9 \sqrt{6} k_3 s_\theta v_{23} - 4 \sqrt{6} m_\pi^2 s_\theta t_1 v_{23} + \sqrt{3} s_\theta t_1 \epsilon\\ & - 2 \sqrt{3} m_K^2 s_\theta t_1 (2 + \sqrt{2} v_{23}) + \sqrt{3} c_\theta t_1 \bigg[-2 m_K^2 (-\sqrt{2} + 2 v_{23}) + m_\pi^2 (4 \sqrt{2} + 7 v_{23}) + 4 \sqrt{2} \epsilon\bigg]\Bigg\} \,,
\end{aligned}
\end{equation}
\begin{equation}
\begin{aligned}
F_{a\gamma\gamma}^{\rm LO}&=\frac{1}{12 F \pi^2}\big[3 v_{41} - 2 \sqrt{6} s_\theta v_{42}  + \sqrt{3} s_\theta v_{43} +\sqrt{3}c_{\theta}(v_{42} + 2 \sqrt{2} v_{43} )\big]\,, \\ 
F_{a\gamma\gamma}^{\rm NLO}&=\frac{1}{12 F \pi^2}\Bigg\{ 3 x_{14} - 2 \sqrt{6} s_\theta x_{24} + 3 v_{12} x_{24}^{(0)} - \sqrt{3} s_\theta v_{23} x_{24}^{(0)}  + \sqrt{3} s_\theta x_{34}+ 3 v_{13} x_{34}^{(0)}\\ & - 2 \sqrt{6} s_\theta v_{23} x_{34}^{(0)} + 3 y_{14} - 2 \sqrt{6} s_\theta y_{24}+ 3 v_{12} y_{24}^{(0)} - \sqrt{3} s_\theta v_{23} y_{24}^{(0)} + \sqrt{3} s_\theta y_{34} \\ &+ 3 v_{13} y_{34}^{(0)} - 2 \sqrt{6} s_\theta v_{23} y_{34}^{(0)} + \sqrt{3} c_\theta \bigg[ x_{24} - 2 \sqrt{2} v_{23} x_{24}^{(0)} + 2 \sqrt{2} x_{34} + v_{23} x_{34}^{(0)}  \\ & + y_{24}- 2 \sqrt{2} v_{23} y_{24}^{(0)} +
2 \sqrt{2} y_{34} + v_{23} y_{34}^{(0)}\bigg]\Bigg\} - \frac{64 k_3 }{3 F} \bigg( - \frac{F}{f_a}-\sqrt{6}s_\theta v_{42}+\sqrt{6}c_\theta v_{43} \bigg)\\ & - \frac{64 t_1 }{27 F } \Bigg\{  \bigg[m_\pi^2  \big[9 v_{41} + \sqrt{3} (7 c_\theta v_{42} - 4 \sqrt{2} s_\theta v_{42} + 4 \sqrt{2} c_\theta v_{43} + 7 s_\theta v_{43}) \big]  + \sqrt{3} \big[c_\theta (v_{42}^{(0)} + 4 \sqrt{2} v_{43}^{(0)}) \epsilon \\ &-2 m_K^2 s_\theta (\sqrt{2} v_{42} + 2 v_{43})+ c_\theta m_K^2 (-4 v_{42} + 2 \sqrt{2} v_{43}) +s_\theta (-4 \sqrt{2} v_{42}^{(0)} + v_{43}^{(0)}) \epsilon \big]\bigg]\Bigg\}\,.
\end{aligned}
\end{equation}
As an improvement of the previous study in Ref.~\cite{Gao:2022xqz}, which only included the terms in the isospin limit, we have now completed the calculation of the linear IB contributions to the two-photon couplings up to NLO in the $\delta$ counting. In principle, the quadratic IB corrections will also appear at NLO. However, as demonstrated later, the linear IB corrections in the NLO part of the two-photon couplings $F_{\phi\gamma\gamma}^{\rm NLO}$ are already tiny for all the cases with $\phi=a, \pi^0, \eta$ and $\eta'$, at the level at most around 1\%, comparing with the results in the isospin limit. Therefore it is expected that the quadratic IB terms can be safely neglected and we will not discuss such effects any more.

Apart from the LECs in the even parity sector that have been fixed via the fit to the lattice data, there are two additional unknown parameters in the WZW Lagrangian, namely $t_1$ and $k_3$ in Eq.~\eqref{eq.lagwzwnloexpd}. These two  parameters can be determined by fitting the experimental decay widths of $\pi^0\to\gamma\gamma$, $\eta\to\gamma\gamma$ and $\eta'\to\gamma\gamma$, which can be equivalently transformed to the experimental inputs of two-photon couplings
\begin{eqnarray}\label{eq.phiggexp}
F_{\pi^0\gamma\gamma}^{\rm Exp} &=& 0.274\pm 0.002 \text{GeV}^{-1} \,, \nonumber\\ 
F_{\eta\gamma\gamma}^{\rm Exp} &=&  0.274\pm 0.006 \text{GeV}^{-1} \,, \nonumber\\ 
F_{\eta'\gamma\gamma}^{\rm Exp} &=& 0.344\pm 0.008 \text{GeV}^{-1} \,,
\end{eqnarray}
where the recent PDG results~\cite{ParticleDataGroup:2022pth} have been used to obtain the above values. 
The fit procedure gives 
\begin{equation}
\begin{aligned}\label{eq.t1k3}
t_1=-(3.8 \pm 2.4)\times 10^{-4}\text{GeV}^{-2},\quad k_3=(1.21 \pm 0.23)\times 10^{-4}
\end{aligned}
\end{equation}
leading to 
\begin{eqnarray}
F_{\pi^0\gamma\gamma}^{\rm Theo}&=& 0.280 \pm 0.001 \text{GeV}^{-1}\,, \\
F_{\eta\gamma\gamma}^{\rm Theo}&=&0.276 \pm 0.009 \text{GeV}^{-1}\,,  \\ 
F_{\eta'\gamma\gamma}^{\rm Theo}&=&0.342 \pm 0.012 \text{GeV}^{-1}\,,
\end{eqnarray}
which are in excellent agreement with the experimental inputs in Eq.~\eqref{eq.phiggexp}. We verify that the IB corrections to $F_{\pi^0\gamma\gamma}, F_{\eta\gamma\gamma}$ and $F_{\eta'\gamma\gamma}$ are small, at the level around $0.5\% \sim 1.5\%$. 

Next we are ready to predict the value of the axion-photon-photon coupling  
\begin{eqnarray}
F_{a\gamma\gamma} = \frac{20.1+3.4+ (0.5 \pm 0.2) }{f_a}  \times 10^{-3} \,,
\end{eqnarray}
where the first term in the numerator denotes the LO contribution in the isospin limit, the second term stands for the linear IB correction at LO, and the third one corresponds to the NLO contribution (including both isospin symmetric and IB effects). In Ref.~\cite{Gao:2022xqz} the error bar of $F_{a\gamma\gamma}$ is purely estimated via the $t_1$ and $k_3$ parameters. While in the present study, we take into account the uncertainties from both the former two parameters and also mixing elements that depend on the parameters of $F,L_5,L_8,\Lambda_1,\Lambda_2$ fitted in Ref.~\cite{Gao:2022xqz}. Comparing with the result in Ref.~\cite{Gao:2022xqz}, it can be inferred that the overwhelmingly dominant IB correction appears in the LO part, while the effect of the IB term in the NLO part is completely negligible. For easy comparison with the determination in literature~\cite{GrillidiCortona:2015jxo,Lu:2020rhp}, we relate the $F_{a\gamma\gamma}$ coupling with $g_{a\gamma\gamma}$ through  
\begin{eqnarray}
g_{a\gamma\gamma} = 4\pi \alpha_{em} F_{a\gamma\gamma}=\frac{\alpha_{em}}{2\pi f_a} \big( 1.89 \pm 0.02 \big) \,. 
\end{eqnarray} 
The updated number inside the bracket is now compatible with the previous result of $1.92\pm0.04$ from the NLO $SU(2)$ $\xpt$~\cite{GrillidiCortona:2015jxo} within uncertainties, and becomes closer to the value of $2.05\pm0.03$ from the NLO $SU(3)$ $\xpt$~\cite{Lu:2020rhp}.

\section{Summary and conclusions}~\label{sec.concl}

In this work we have calculated the mixing of the $\pi^0$-$\eta$-$\eta'$-axion system and their two-photon couplings within the $U(3)$ chiral perturbation theory up to next-to-leading order by including the complete linear strong isospin-breaking terms, i.e. the correction  proportional to $m_u-m_d$. The isospin-breaking correction constitutes the key improvement of this work, comparing with the previous calculation in Ref.~\cite{Gao:2022xqz}. 
The quantities considered in the fits to the lattice data in the former reference, including the $m_\pi$ dependence of $m_\eta, m_{\eta'},m_K,F_\pi,F_K$ and the $\eta$-$\eta'$ mixing parameters, are found to be barely affected after including the full linear isospin-breaking corrections.  

For the two-photon couplings of $\pi^0$, $\eta$, $\eta'$ and axion, the isospin-breaking corrections to the former three light pseudoscalar mesons are quite small. While, for the axion-photon-photon coupling, it receives quite important contributions from the isospin-breaking terms, acquiring more than 15\% correction comparing with the isospin symmetric part. The isospin-breaking correction to the axion two-photon coupling in $U(3)$ chiral theory renders its prediction roughly compatible with those in the $SU(2)$ and $SU(3)$ cases.

\section*{Acknowledgment}
We would like to thank Luca Di Luzio for interesting and inspiring discussions on $g_{a\gamma\gamma}$. 
This work is funded in part by the National Natural Science Foundation of China (NSFC) under Grants Nos.~12150013, 12475078, 11975090, and the Science Foundation of Hebei Normal University with contract No.~L2023B09. JAO would like to acknowledge partial financial support to the Grant PID2022-136510NB-C32 funded by MCIN/AEI/10.13039/501100011033/ and FEDER, UE, and to the EU Horizon 2020 research and innovation program, STRONG-2020 project, under grant agreement no. 824093. 

\section*{Appendix: Next-to-leading order coefficients for the bilinear terms in Eq.~\eqref{eq.lagsenlo}}

Here we provide the lengthy expressions of the $\delta_i$ appearing in the general bilinear terms in Eq.~\eqref{eq.lagsenlo} by including the complete linear IB corrections. Although some of the formulas are the same as those in Ref.~\cite{Gao:2022xqz}, we give all the relevant expressions below for the sake of completeness:  

\begin{equation}\label{eq.begindiff}
\begin{aligned}
\delta^{a\pi}_{k}=&\frac{8 L_5}{3 F^2} \bigg\{2 c_\theta^2 m_K^2 (-2 v_{12} v_{42}^{(0)} + \sqrt{2} v_{13} v_{42}^{(0)} + \sqrt{2} v_{12} v_{43}^{(0)} - v_{13} v_{43}^{(0)})\\ & + m_\pi^2 \bigg[3 v_{41} + s_\theta^2 (-v_{12} v_{42}^{(0)} + 2 \sqrt{2} v_{13} v_{42}^{(0)} + 2 \sqrt{2} v_{12} v_{43}^{(0)} + v_{13} v_{43}^{(0)}) + c_\theta^2 \big[-v_{13} (2 \sqrt{2} v_{42}^{(0)} + v_{43}^{(0)})\\ & + v_{12} (v_{42}^{(0)} - 2 \sqrt{2} v_{43}^{(0)})\big] + 2 c_\theta s_\theta \big[v_{12} (2 \sqrt{2} v_{42}^{(0)} + v_{43}^{(0)}) + v_{13} (v_{42}^{(0)} - 2 \sqrt{2} v_{43}^{(0)})\big]\bigg]\\ & - s_\theta \bigg[2 m_K^2 s_\theta (v_{12} v_{42}^{(0)} + \sqrt{2} v_{13} v_{42}^{(0)} + \sqrt{2} v_{12} v_{43}^{(0)} + 2 v_{13} v_{43}^{(0)}) + \sqrt{3} (\sqrt{2} v_{42}^{(0)} - v_{43}^{(0)}) \epsilon\bigg] \\ &+ c_\theta \bigg[-2 m_K^2 s_\theta \big[v_{12} (2 \sqrt{2} v_{42}^{(0)} + v_{43}^{(0)}) + v_{13} (v_{42}^{(0)} - 2 \sqrt{2} v_{43}^{(0)})\big] + \sqrt{3} (v_{42}^{(0)} + \sqrt{2} v_{43}^{(0)}) \epsilon\bigg]\bigg\} \\ & - \frac{1}{6 f_a}\bigg[(s_\theta v_{12} - c_\theta v_{13}) (\sqrt{6} F + 6 f_a s_\theta v_{42}^{(0)} - 6 c_\theta f_a v_{43}^{(0)})\Lambda_1\bigg],
\end{aligned}
\end{equation}
\begin{equation}
\begin{aligned}
\delta^{a\eta}_{k}=&\frac{8 L_5}{3 F^2} \bigg\{c_\theta s_\theta \bigg[2 m_K^2 (2 \sqrt{2} v_{42} - v_{23} v_{42}^{(0)} + v_{43} + 2 \sqrt{2} v_{23} v_{43}^{(0)}) \\ &- 2 m_\pi^2 (2 \sqrt{2} v_{42} - v_{23} v_{42}^{(0)} + v_{43} + 2 \sqrt{2} v_{23} v_{43}^{(0)}) + (2 \sqrt{2} v_{42}^{(0)} + v_{43}^{(0)}) \epsilon \bigg]\\ & + c_\theta^2 \bigg[-m_\pi^2 (v_{42} + 2 \sqrt{2} v_{23} v_{42}^{(0)} - 2 \sqrt{2} v_{43} + v_{23} v_{43}^{(0)}) + 2 m_K^2 \big[(2 v_{42} + \sqrt{2} v_{23}v_{42}^{(0)})\\ &  - (\sqrt{2}v_{43} + v_{23}v_{43}^{(0)}) \big] + (2 v_{42}^{(0)} - \sqrt{2} v_{43}^{(0)}) \epsilon\bigg] + s_\theta^2 \bigg[2 m_K^2 \big[v_{42} - \sqrt{2} v_{23} v_{42}^{(0)} \\ &+(\sqrt{2}v_{43} - 2 v_{23}v_{43}^{(0)}) \big] + m_\pi^2 (v_{42} + 2 \sqrt{2} v_{23} v_{42}^{(0)} - 2 \sqrt{2} v_{43}+ v_{23} v_{43}^{(0)}) + (v_{42}^{(0)} + \sqrt{2} v_{43}^{(0)}) \epsilon\bigg]\bigg\} \\ &+\frac{1}{6 f_a} \bigg[(s_\theta + c_\theta v_{23}) (\sqrt{6} F + 6 f_a s_\theta v_{42}^{(0)} - 6 c_\theta f_a v_{43}^{(0)}) \Lambda_1+s_\theta( 6 f_a s_\theta v_{42}^{(1)} - 6 c_\theta f_a v_{43}^{(1)})\Lambda_1\bigg],
\end{aligned}
\end{equation}
\begin{equation}
\begin{aligned}
\delta^{a\eta'}_{k}=&\frac{8 L_5}{3 F^2}\bigg\{c_\theta^2 \bigg[m_\pi^2 (2 \sqrt{2} v_{42} - v_{23} v_{42}^{(0)} + v_{43}+ 2 \sqrt{2} v_{23} v_{43}^{(0)}) \\ &- 
2 m_K^2 \big[(\sqrt{2}v_{42} - 2 v_{23}v_{42}^{(0)})  + (-v_{43} + \sqrt{2} v_{23}v_{43}^{(0)}) \big] \\ &+ (-\sqrt{2} v_{42}^{(0)} + v_{43}^{(0)}) \epsilon\bigg] +s_\theta^2 \bigg[-m_\pi^2 (2 \sqrt{2} v_{42} - v_{23} v_{42}^{(0)} + v_{43}+ 2 \sqrt{2} v_{23} v_{43}^{(0)})\\ & + 
2 m_K^2 \big[(\sqrt{2}v_{42} + v_{23}v_{42}^{(0)})  + (2 v_{43}+ \sqrt{2} v_{23}v_{43}^{(0)}) \big] + (\sqrt{2} v_{42}^{(0)} + 2 v_{43}^{(0)}) \epsilon\bigg]\\ & + 
c_\theta s_\theta \bigg[-2 m_\pi^2 (v_{42} + 2 \sqrt{2} v_{23} v_{42}^{(0)} - 2 \sqrt{2} v_{43} + v_{23} v_{43}^{(0)}) \\ &+ 2 m_K^2 \big[ v_{42} + 2 \sqrt{2} v_{23} v_{42}^{(0)} + (-2 \sqrt{2}v_{43} + v_{23}v_{43}^{(0)}) \big] + (v_{42}^{(0)} - 2 \sqrt{2} v_{43}^{(0)}) \epsilon\bigg]\bigg\}\\ &-\frac{1}{6 f_a} \bigg[(c_\theta - s_\theta v_{23}) (\sqrt{6} F + 6 f_a s_\theta v_{42}^{(0)} - 6 c_\theta f_a v_{43}^{(0)}) \Lambda_1+c_\theta ( 6 f_a s_\theta v_{42}^{(1)} - 6 c_\theta f_a v_{43}^{(1)})\Lambda_1\bigg],
\end{aligned}
\end{equation}
\begin{equation}
\begin{aligned}
\delta^{a}_{k}=&\frac{8 L_5}{3 F^2} \bigg\{4 c_\theta (m_K^2 - m_\pi^2) s_\theta (\sqrt{2} v_{42}^{(0)2} + v_{42}^{(0)} v_{43}^{(0)} - \sqrt{2} v_{43}^{(0)2}) + c_\theta^2 \bigg[2 m_K^2 (2 v_{42}^{(0)2} - 2 \sqrt{2} v_{42}^{(0)} v_{43}^{(0)} + v_{43}^{(0)2}) \\ &+m_\pi^2 (-v_{42}^{(0)2} + 4 \sqrt{2} v_{42}^{(0)} v_{43}^{(0)} + v_{43}^{(0)2})\bigg] + s_\theta^2 \bigg[m_\pi^2 (v_{42}^{(0)2} - 4 \sqrt{2} v_{42}^{(0)} v_{43}^{(0)} - v_{43}^{(0)2}) \\ &+ 2 m_K^2 (v_{42}^{(0)2} + 2 \sqrt{2} v_{42}^{(0)} v_{43}^{(0)} + 2 v_{43}^{(0)2})\bigg]\bigg\} +\frac{\Lambda_1}{6 f_a^2}\bigg\{F^2 + 6 f_a^2 (s_\theta v_{42}^{(0)} - c_\theta v_{43}^{(0)})^2 \\ &- 2 \sqrt{6}F f_a (-s_\theta v_{42}^{(0)} + c_\theta v_{43}^{(0)})\bigg\} +\frac{8 L_5}{3 F^2} \bigg\{c_\theta^2 \bigg[4 m_K^2 (2 v_{42}^{(0)} v_{42}^{(1)} - \sqrt{2} v_{42}^{(1)} v_{43}^{(0)} - \sqrt{2} v_{42}^{(0)} v_{43}^{(1)} + v_{43}^{(0)} v_{43}^{(1)}) \\ &+ m_\pi^2 (-2 v_{42}^{(0)} v_{42}^{(1)} + 4 \sqrt{2} v_{42}^{(1)} v_{43}^{(0)} + 4 \sqrt{2} v_{42}^{(0)} v_{43}^{(1)} + 2 v_{43}^{(0)} v_{43}^{(1)}) + (2 v_{42}^{(0)2} - 2 \sqrt{2} v_{42}^{(0)} v_{43}^{(0)} + v_{43}^{(0)2}) \epsilon\bigg]\\ & + s_\theta^2 \bigg[-2 m_\pi^2 (-v_{42}^{(0)} v_{42}^{(1)} + 2 \sqrt{2} v_{42}^{(1)} v_{43}^{(0)} + 2 \sqrt{2} v_{42}^{(0)} v_{43}^{(1)} + v_{43}^{(0)} v_{43}^{(1)})\\ & + 4 m_K^2 (v_{42}^{(0)} v_{42}^{(1)} + \sqrt{2} v_{42}^{(1)} v_{43}^{(0)} + \sqrt{2} v_{42}^{(0)} v_{43}^{(1)} + 2 v_{43}^{(0)} v_{43}^{(1)})  + (v_{42}^{(0)2} + 2 \sqrt{2} v_{42}^{(0)} v_{43}^{(0)} + 2 v_{43}^{(0)2}) \epsilon\bigg] \\ &+ 2 c_\theta s_\theta \bigg[-2 m_\pi^2 (2 \sqrt{2} v_{42}^{(0)} v_{42}^{(1)} + v_{42}^{(1)} v_{43}^{(0)} + v_{42}^{(0)} v_{43}^{(1)} - 2 \sqrt{2} v_{43}^{(0)} v_{43}^{(1)}) \\ & + 2 m_K^2 \big[v_{42}^{(0)} (2 \sqrt{2} v_{42}^{(1)} + v_{43}^{(1)}) + v_{43}^{(0)} (v_{42}^{(1)} - 2 \sqrt{2} v_{43}^{(1)})\big] + (\sqrt{2} v_{42}^{(0)2} + v_{42}^{(0)} v_{43}^{(0)} - \sqrt{2} v_{43}^{(0)2}) \epsilon\bigg] \bigg\}\\ &+\frac{ \Lambda_1}{3 f_a}\bigg\{(\sqrt{6} F + 6 f_a s_\theta v_{42}^{(0)} - 6 c_\theta f_a v_{43}^{(0)}) (s_\theta v_{42}^{(1)} - c_\theta v_{43}^{(1)})\bigg\},
\end{aligned}
\end{equation}
\begin{equation}
\begin{aligned}
\delta^{a\pi}_{m^{2}}=&\frac{16 L_8}{3 F^2}\bigg\{3 m_\pi^4 \bigg[v_{41} - (v_{12} v_{42}^{(0)} + v_{13} v_{43}^{(0)})\bigg] - 4 m_K^4 \bigg[c_\theta^2 (2 v_{12} v_{42}^{(0)} - \sqrt{2} v_{13} v_{42}^{(0)} - \sqrt{2} v_{12} v_{43}^{(0)} + v_{13} v_{43}^{(0)}) \\ &+ s_\theta^2 (v_{12} v_{42}^{(0)} + \sqrt{2} v_{13} v_{42}^{(0)} + \sqrt{2} v_{12} v_{43}^{(0)} + 2 v_{13} v_{43}^{(0)}) + c_\theta s_\theta (2 \sqrt{2} v_{12} v_{42}^{(0)} + v_{13} v_{42}^{(0)} + v_{12} v_{43}^{(0)} \\ &- 2 \sqrt{2} v_{13} v_{43}^{(0)})\bigg] + 2 m_\pi^2 \bigg[2 c_\theta^2 m_K^2 (2 v_{12} v_{42}^{(0)} - \sqrt{2} v_{13} v_{42}^{(0)} - \sqrt{2} v_{12} v_{43}^{(0)} + v_{13} v_{43}^{(0)}) \\ &+ 2 m_K^2 s_\theta^2 (v_{12} v_{42}^{(0)} + \sqrt{2} v_{13} v_{42}^{(0)} + \sqrt{2} v_{12} v_{43}^{(0)} + 2 v_{13} v_{43}^{(0)}) + 2 c_\theta m_K^2 s_\theta \big[v_{12} (2 \sqrt{2} v_{42}^{(0)} + v_{43}^{(0)})\\ & + v_{13} (v_{42}^{(0)} - 2 \sqrt{2} v_{43}^{(0)})\big] + \sqrt{3} s_\theta (-\sqrt{2} v_{42}^{(0)} + v_{43}^{(0)}) \epsilon + \sqrt{3} c_\theta (v_{42}^{(0)}+ \sqrt{2} v_{43}^{(0)}) \epsilon\bigg]\bigg\} \\ &- \frac{1}{18 f_a}\bigg\{- 12 c_\theta^2 f_a \bigg[m_K^2 (\sqrt{2} v_{13} v_{42}^{(0)} + \sqrt{2} v_{12} v_{43}^{(0)} - 2 v_{13} v_{43}^{(0)}) - m_\pi^2 \big[\sqrt{2} v_{12} v_{43}^{(0)} \\ & + v_{13} (\sqrt{2} v_{42}^{(0)} + v_{43}^{(0)})\big]\bigg] + F \big[\sqrt{3} m_\pi^2 s_\theta (\sqrt{2} v_{12} - 4 v_{13}) + 2 \sqrt{3} m_K^2 s_\theta (\sqrt{2} v_{12} + 2 v_{13}) + 6 \epsilon\big] \\ & + 6 f_a s_\theta \big[-2 m_\pi^2 s_\theta (-v_{12} v_{42}^{(0)} + \sqrt{2} v_{13} v_{42}^{(0)} +\sqrt{2} v_{12} v_{43}^{(0)}) + 2 m_K^2 s_\theta (2 v_{12} v_{42}^{(0)}\\ & + \sqrt{2} v_{13} v_{42}^{(0)} + \sqrt{2} v_{12} v_{43}^{(0)}) + \sqrt{6} v_{42}^{(0)} \epsilon\big] + c_\theta \bigg[\sqrt{3}F \big[m_K^2 (4 v_{12} - 2 \sqrt{2} v_{13}) - m_\pi^2 (4 v_{12} + \sqrt{2} v_{13})\big] \\ &- 6 f_a \big[2 m_\pi^2 s_\theta (2 \sqrt{2} v_{12} v_{42}^{(0)} + v_{13} v_{42}^{(0)} + v_{12} v_{43}^{(0)} - 2 \sqrt{2} v_{13} v_{43}^{(0)}) \\ &+ 4 m_K^2 s_\theta \big[v_{12} (-\sqrt{2} v_{42}^{(0)} + v_{43}^{(0)}) + v_{13} (v_{42}^{(0)} + \sqrt{2} v_{43}^{(0)})\big] + \sqrt{6} v_{43}^{(0)} \epsilon\big]\bigg]\bigg\} \Lambda_2,
\end{aligned}
\end{equation}
\begin{equation}
\begin{aligned}
\delta^{a\eta}_{m^{2}}=&\frac{16 L_8}{3 F^2} \bigg\{c_\theta^2 \bigg[4 m_K^4 \big[(2 v_{42} + \sqrt{2} v_{23} v_{42}^{(0)}) - (\sqrt{2}v_{43} + v_{23}v_{43}^{(0)}) \big] - 4 m_K^2 \big[m_\pi^2 ((2v_{42} + \sqrt{2} v_{23}v_{42}^{(0)}) \\ &-(\sqrt{2}v_{43} + v_{23}v_{43}^{(0)})) - 2 v_{42}^{(0)} \epsilon + \sqrt{2} v_{43}^{(0)} \epsilon\big] + 
m_\pi^2 \big[3 m_\pi^2 (v_{42} - v_{23} v_{43}^{(0)}) - 4 v_{42}^{(0)} \epsilon + 2 \sqrt{2} v_{43}^{(0)} \epsilon\big]\bigg] \\[-0.3em] &+ 2 c_\theta s_\theta \bigg[2 m_K^4 (2 \sqrt{2} v_{42} - v_{23} v_{42}^{(0)} + v_{43} + 2 \sqrt{2} v_{23} v_{43}^{(0)}) - m_\pi^2 (2 \sqrt{2} v_{42}^{(0)} + v_{43}^{(0)}) \epsilon \\[-0.3em] &+ m_K^2 \big[-2 m_\pi^2 (2 \sqrt{2} v_{42}  - v_{23} v_{42}^{(0)} + v_{43} + 2 \sqrt{2} v_{23} v_{43}^{(0)}) + 2 (2 \sqrt{2} v_{42}^{(0)} + v_{43}^{(0)}) \epsilon\big]\bigg] \\[-0.3em] &+ s_\theta^2 \bigg[4 m_K^4 \big[v_{42} - \sqrt{2} v_{23} v_{42}^{(0)} + (\sqrt{2}v_{43}- 2 v_{23}v_{43}^{(0)}) \big] + 3 m_\pi^4 (v_{42} - v_{23} v_{43}^{(0)}) - 2 m_\pi^2 (v_{42}^{(0)} + \sqrt{2} v_{43}^{(0)}) \epsilon \\[-0.3em] & + 4 m_K^2 \big[m_\pi^2 (-v_{42} + \sqrt{2} v_{23} v_{42}^{(0)} - \sqrt{2} v_{43}+ 2 v_{23} v_{43}^{(0)}) + (v_{42}^{(0)} + \sqrt{2} v_{43}^{(0)}) \epsilon\big]\bigg]\bigg\}  \\[-0.3em] & + \frac{1}{18 f_a}\bigg\{- 6 c_\theta^2 f_a \bigg[2 m_\pi^2 (\sqrt{2} v_{23} v_{42}^{(0)} - \sqrt{2} v_{43} + v_{23} v_{43}^{(0)}) + m_K^2 (-2 \sqrt{2} v_{23} v_{42}^{(0)} \\[-0.3em] &+ 2 \sqrt{2} v_{43} + 4 v_{23} v_{43}^{(0)}) + \sqrt{2} v_{43}^{(0)} \epsilon\bigg] + c_\theta \bigg[\sqrt{3}F (m_\pi^2 (-4 + \sqrt{2} v_{23}) + 2 m_K^2 (2 + \sqrt{2} v_{23}) + 2 \epsilon) \\ & + 12 f_a s_\theta \big[-m_\pi^2 (2 \sqrt{2} v_{42} - v_{23} v_{42}^{(0)} + v_{43} + 2 \sqrt{2} v_{23} v_{43}^{(0)}) + 2 m_K^2 \big[(\sqrt{2}v_{42} + v_{23}v_{42}^{(0)}) \\[-0.3em] & + (-v_{43} + \sqrt{2} v_{23}v_{43}^{(0)}) \big] + (\sqrt{2} v_{42}^{(0)} - v_{43}^{(0)}) \epsilon\big]\bigg] + s_\theta \bigg[\sqrt{3}F \big[2 m_K^2 (\sqrt{2} - 2 v_{23}) \\ &+ m_\pi^2 (\sqrt{2} + 4 v_{23}) + \sqrt{2} \epsilon \big] + 6 f_a s_\theta \big[2 m_\pi^2 (v_{42} + \sqrt{2} v_{23} v_{42}^{(0)} - \sqrt{2} v_{43}) \\[-0.3em] &+ m_K^2 \big[(4v_{42} - 2 \sqrt{2} v_{23}v_{42}^{(0)})  + 2 \sqrt{2} v_{43}\big] + (2 v_{42}^{(0)} +\sqrt{2} v_{43}^{(0)}) \epsilon\big]\bigg]\bigg\} \Lambda_2,
\end{aligned}
\end{equation}
\begin{equation}
\begin{aligned}
\delta_{m_a^{2}}=&\frac{16 L_8}{3 F^2}\bigg\{8 c_\theta m_K^2 (m_K^2 - m_\pi^2) s_\theta (\sqrt{2} v_{42}^{(0)2} + v_{42}^{(0)} v_{43}^{(0)} - \sqrt{2} v_{43}^{(0)2}) + c_\theta^2 \bigg[3 m_\pi^4 (v_{42}^{(0)2} + v_{43}^{(0)2}) \\ &+ 
4 m_K^4 (2 v_{42}^{(0)2} - 2 \sqrt{2} v_{42}^{(0)} v_{43}^{(0)} + v_{43}^{(0)2}) - 4 m_K^2 m_\pi^2 (2 v_{42}^{(0)2} - 2 \sqrt{2} v_{42}^{(0)} v_{43}^{(0)} + v_{43}^{(0)2})\bigg]\\ & + 
s_\theta^2 \bigg[3 m_\pi^4 (v_{42}^{(0)2} + v_{43}^{(0)2}) + 
4 m_K^4 (v_{42}^{(0)2} + 2 \sqrt{2} v_{42}^{(0)} v_{43}^{(0)} + 2 v_{43}^{(0)2}) - 4 m_K^2 m_\pi^2 (v_{42}^{(0)2} + 2 \sqrt{2} v_{42}^{(0)} v_{43}^{(0)}\\ & + 2 v_{43}^{(0)2})\bigg]\bigg\} - 
\frac{\Lambda_2}{9 f_a}\bigg\{-6 c_\theta^2 f_a v_{43}^{(0)} \bigg[m_\pi^2 (2 \sqrt{2} v_{42}^{(0)} + v_{43}^{(0)}) + m_K^2 (-2 \sqrt{2} v_{42}^{(0)} + 2 v_{43}^{(0)})\bigg] \\ &+ s_\theta \bigg[-\sqrt{3} F \big[m_\pi^2 (\sqrt{2} v_{42}^{(0)} - 4 v_{43}^{(0)}) + 2 m_K^2 (\sqrt{2} v_{42}^{(0)} + 2 v_{43}^{(0)})\big] - 6 f_a s_\theta v_{42}^{(0)} \big[m_\pi^2 (v_{42}^{(0)} - 2 \sqrt{2} v_{43}^{(0)}) \\ &+ 2 m_K^2 (v_{42}^{(0)} + \sqrt{2} v_{43}^{(0)})\big]\bigg] + c_\theta \bigg[-12 f_a s_\theta \big[m_K^2 (\sqrt{2} v_{42}^{(0)2} - 2 v_{42}^{(0)} v_{43}^{(0)} -\sqrt{2} v_{43}^{(0)2}) - m_\pi^2 (\sqrt{2} v_{42}^{(0)2} \\ & + v_{42}^{(0)} v_{43}^{(0)} - \sqrt{2} v_{43}^{(0)2})\big] + \sqrt{3} F \big[m_\pi^2 (4 v_{42}^{(0)} + \sqrt{2} v_{43}^{(0)}) + m_K^2 (-4 v_{42}^{(0)} + 2 \sqrt{2} v_{43}^{(0)}) - 2 v_{42}^{(0)} \epsilon\big]\bigg]\bigg\} \\ &+\frac{32 L_8}{3 F^2} \bigg\{c_\theta^2 \bigg[3 m_\pi^4 (v_{42}^{(0)} v_{42}^{(1)} + v_{43}^{(0)} v_{43}^{(1)}) + 4 m_K^4 (2 v_{42}^{(0)} v_{42}^{(1)} - \sqrt{2} v_{42}^{(1)} v_{43}^{(0)} - \sqrt{2} v_{42}^{(0)} v_{43}^{(1)} + v_{43}^{(0)} v_{43}^{(1)}) \\ &- m_\pi^2 (2 v_{42}^{(0)2} - 2 \sqrt{2} v_{42}^{(0)} v_{43}^{(0)} + v_{43}^{(0)2}) \epsilon +m_K^2 \big[m_\pi^2 (-8 v_{42}^{(0)} v_{42}^{(1)} + 4 \sqrt{2} v_{42}^{(1)} v_{43}^{(0)} + 4 \sqrt{2} v_{42}^{(0)} v_{43}^{(1)} \\ &- 4 v_{43}^{(0)} v_{43}^{(1)}) + 2 (2 v_{42}^{(0)2} - 2 \sqrt{2} v_{42}^{(0)} v_{43}^{(0)} + 
v_{43}^{(0)2}) \epsilon\big]\bigg] + s_\theta^2 \bigg[3 m_\pi^4 (v_{42}^{(0)} v_{42}^{(1)} + v_{43}^{(0)} v_{43}^{(1)}) \\ &+ 4 m_K^4 (v_{42}^{(0)} v_{42}^{(1)} + \sqrt{2} v_{42}^{(1)} v_{43}^{(0)} + \sqrt{2} v_{42}^{(0)} v_{43}^{(1)} + 2 v_{43}^{(0)} v_{43}^{(1)}) - 
m_\pi^2 (v_{42}^{(0)2} + 2 \sqrt{2} v_{42}^{(0)} v_{43}^{(0)} + 2 v_{43}^{(0)2}) \epsilon\\ & +m_K^2 \big[-4 m_\pi^2 (v_{42}^{(0)} v_{42}^{(1)} + \sqrt{2} v_{42}^{(1)} v_{43}^{(0)} + 
\sqrt{2} v_{42}^{(0)} v_{43}^{(1)} + 2 v_{43}^{(0)} v_{43}^{(1)}) \\ &+ 2 (v_{42}^{(0)2} + 2 \sqrt{2} v_{42}^{(0)} v_{43}^{(0)} + 2 v_{43}^{(0)2}) \epsilon\big]\bigg]  + 2 c_\theta s_\theta \bigg[2 m_K^4 \big[v_{42}^{(0)} (2 \sqrt{2} v_{42}^{(1)} + v_{43}^{(1)}) \\ &+ v_{43}^{(0)} (v_{42}^{(1)} - 2 \sqrt{2} v_{43}^{(1)})\big] - m_\pi^2 (\sqrt{2} v_{42}^{(0)2} + v_{42}^{(0)} v_{43}^{(0)} - \sqrt{2} v_{43}^{(0)2}) \epsilon- 2 m_K^2 \big[m_\pi^2 (2 \sqrt{2} v_{42}^{(0)} v_{42}^{(1)}\\ & + v_{42}^{(1)} v_{43}^{(0)} + 
v_{42}^{(0)} v_{43}^{(1)} - 2 \sqrt{2} v_{43}^{(0)} v_{43}^{(1)}) - (\sqrt{2} v_{42}^{(0)2} + v_{42}^{(0)} v_{43}^{(0)} - \sqrt{2} v_{43}^{(0)2}) \epsilon\big]\bigg]\bigg\}\\ &+\frac{ \Lambda_2}{9 f_a}\bigg\{-6 c_\theta^2 f_a \bigg[2 m_K^2 (\sqrt{2} v_{42}^{(1)} v_{43}^{(0)} + \sqrt{2} v_{42}^{(0)} v_{43}^{(1)} - 2 v_{43}^{(0)} v_{43}^{(1)}) - 2 m_\pi^2 (\sqrt{2} v_{42}^{(1)} v_{43}^{(0)} \\ & + \sqrt{2} v_{42}^{(0)} v_{43}^{(1)} + v_{43}^{(0)} v_{43}^{(1)}) + (\sqrt{2} v_{42}^{(0)} - v_{43}^{(0)}) v_{43}^{(0)} \epsilon\bigg]  + s_\theta \bigg[\sqrt{3}F \big[m_\pi^2 (\sqrt{2} v_{42}^{(1)} - 4 v_{43}^{(1)}) \\ & + 2 m_K^2 (\sqrt{2} v_{42}^{(1)} + 2 v_{43}^{(1)}) + (\sqrt{2} v_{42}^{(0)} + 2 v_{43}^{(0)}) \epsilon\big] + 6 f_a s_\theta \big[-2 m_\pi^2 (-v_{42}^{(0)} v_{42}^{(1)}\\ &  + \sqrt{2} v_{42}^{(1)} v_{43}^{(0)} + \sqrt{2} v_{42}^{(0)} v_{43}^{(1)})  + 2 m_K^2 (2 v_{42}^{(0)} v_{42}^{(1)} + \sqrt{2} v_{42}^{(1)} v_{43}^{(0)} + \sqrt{2} v_{42}^{(0)} v_{43}^{(1)}) \\ &+ v_{42}^{(0)} (v_{42}^{(0)} + \sqrt{2} v_{43}^{(0)}) \epsilon\big]\bigg] + c_\theta \bigg[-\sqrt{3}F \big[m_\pi^2 (4 v_{42}^{(1)} + \sqrt{2} v_{43}^{(1)}) + m_K^2 (-4 v_{42}^{(1)} + 2 \sqrt{2} v_{43}^{(1)})  \\ & - 2 v_{42}^{(1)} \epsilon + \sqrt{2} v_{43}^{(0)} \epsilon\big] + 6 f_a s_\theta \big[-2 m_\pi^2 (2 \sqrt{2} v_{42}^{(0)} v_{42}^{(1)} + v_{42}^{(1)} v_{43}^{(0)} +v_{42}^{(0)} v_{43}^{(1)} - 2 \sqrt{2} v_{43}^{(0)} v_{43}^{(1)}) \\ &+ 4 m_K^2 (v_{42}^{(0)} (\sqrt{2} v_{42}^{(1)} - v_{43}^{(1)}) - v_{43}^{(0)} (v_{42}^{(1)} + \sqrt{2} v_{43}^{(1)})) + (\sqrt{2} v_{42}^{(0)2} - 2 v_{42}^{(0)} v_{43}^{(0)} - \sqrt{2} v_{43}^{(0)2}) \epsilon\big]\bigg]\bigg\},
\end{aligned}
\end{equation}
\begin{equation}
\begin{aligned}
\delta^{a\eta'}_{m^{2}}=&\frac{16 L_8}{3 F^2}
 \bigg\{c_\theta^2 \bigg[3 m_\pi^4 (v_{23} v_{42}^{(0)} + v_{43}) - 4 m_K^4 \big[(\sqrt{2}v_{42} - 2 v_{23}v_{42}^{(0)})  + (- v_{43}+ \sqrt{2} v_{23} v_{43}^{(0)})\big] \\ &+ 
2 m_\pi^2 (\sqrt{2} v_{42}^{(0)} - v_{43}^{(0)}) \epsilon + 4 m_K^2 \big[m_\pi^2 (\sqrt{2} v_{42} - 2 v_{23} v_{42}^{(0)}- v_{43} + \sqrt{2} v_{23} v_{43}^{(0)}) + (-\sqrt{2} v_{42}^{(0)} + v_{43}^{(0)}) \epsilon\big]\bigg]\\ & +
s_\theta^2 \bigg[3 m_\pi^4 (v_{23} v_{42}^{(0)} + v_{43}) + 4 m_K^4 \big[(\sqrt{2}v_{42} + v_{23}v_{42}^{(0)})  + (2v_{43} + \sqrt{2} v_{23}v_{43}^{(0)}) \big]  \\ &- 2 m_\pi^2 (\sqrt{2} v_{42}^{(0)} + 2 v_{43}^{(0)}) \epsilon- 4 m_K^2 \big[m_\pi^2 (\sqrt{2}v_{42} + v_{23}v_{42}^{(0)}  + 2 v_{43} + \sqrt{2} v_{23}v_{43}^{(0)} ) - (\sqrt{2} v_{42}^{(0)} + 2 v_{43}^{(0)}) \epsilon\big]\bigg] \\ &+2 c_\theta s_\theta \bigg[2 m_K^4 \big[v_{42} + 2 \sqrt{2} v_{23} v_{42}^{(0)} + (-2 \sqrt{2} v_{43}+ v_{23}v_{43}^{(0)})  \big]- m_\pi^2 (v_{42}^{(0)} - 2 \sqrt{2} v_{43}^{(0)}) \epsilon \\ &- 2 m_K^2 \big[m_\pi^2 (v_{42} + 2 \sqrt{2} v_{23} v_{42}^{(0)} - 2 \sqrt{2} v_{43} + v_{23} v_{43}^{(0)}) - (v_{42}^{(0)} - 2 \sqrt{2} v_{43}^{(0)}) \epsilon\big]\bigg]\bigg\} \\ & - \frac{1}{18 f_a}\bigg\{6 c_\theta^2 f_a \bigg[-2 m_\pi^2 (\sqrt{2} v_{42} + v_{43}  + \sqrt{2} v_{23} v_{43}^{(0)}) + 2 m_K^2 \big[\sqrt{2} v_{42} + (-2 v_{43}+ \sqrt{2} v_{23}v_{43}^{(0)}) \big]\\ & + (\sqrt{2} v_{42}^{(0)} - 2 v_{43}^{(0)}) \epsilon\bigg] + s_\theta \bigg[-\sqrt{3} F \big[m_\pi^2 (-4 + \sqrt{2} v_{23}) + 2 m_K^2 (2 + \sqrt{2} v_{23}) + 2 \epsilon \big] \\ & - 6 f_a s_\theta \big[-2 m_\pi^2 (\sqrt{2} v_{42} +v_{23}(- v_{42}^{(0)} + \sqrt{2} v_{43}^{(0)})) + 2 m_K^2 (\sqrt{2}v_{42} +v_{23} ( 2 v_{42}^{(0)}  + \sqrt{2}  v_{43}^{(0)})) + \sqrt{2} v_{42}^{(0)} \epsilon\big]\bigg] \\ &+ c_\theta \bigg[\sqrt{3} F \big[2 m_K^2 (\sqrt{2} - 2 v_{23}) + m_\pi^2 (\sqrt{2} + 4 v_{23}) + \sqrt{2} \epsilon\big] - 12 f_a s_\theta \big[-m_\pi^2 (v_{42} + 2 \sqrt{2} v_{23} v_{42}^{(0)} \\ &- 2 \sqrt{2} v_{43}+ v_{23} v_{43}^{(0)}) + 2 m_K^2 \big[(-v_{42} + \sqrt{2} v_{23}v_{42}^{(0)})  - (\sqrt{2} v_{43} + v_{23} v_{43}^{(0)})\big] - (v_{42}^{(0)} + \sqrt{2} v_{43}^{(0)}) \epsilon\big]\bigg]\bigg\} \Lambda_2,
\end{aligned}
\end{equation}

\begin{equation}
\begin{aligned}
\delta^{\eta \eta'}_{k}=&\frac{8 L_5}{3 F^2}
\Bigg\{c_\theta s_\theta \Bigg[-2 m_\pi^2 (1 + 4 \sqrt{2} v_{23}) + m_K^2 (2 + 8 \sqrt{2} v_{23}) + \epsilon \Bigg]\\ & - 
c_\theta^2 \Bigg[2 m_K^2 (\sqrt{2} - v_{23}) - 2 m_\pi^2 (\sqrt{2} - v_{23}) + \sqrt{2} \epsilon \Bigg]\\ & + 
s_\theta^2 \Bigg[2 m_K^2 (\sqrt{2} - v_{23}) - 2 m_\pi^2 (\sqrt{2} - v_{23}) + \sqrt{2} \epsilon\Bigg]\Bigg\} - (c_\theta s_\theta + c_\theta^2 v_{23} - s_\theta^2 v_{23}) \Lambda_1,
\end{aligned}
\end{equation}
\begin{equation}
\begin{aligned}
\delta^{\eta}_{k}=&\frac{8 L_5}{3 F^2}\Bigg\{s_\theta^2 \Bigg[m_K^2 (2 - 4 \sqrt{2} v_{23}) + m_\pi^2 (1 + 4 \sqrt{2} v_{23}) + \epsilon\Bigg] \\ &+ c_\theta^2 \Bigg[4 m_K^2 (1 + \sqrt{2} v_{23}) - m_\pi^2 (1 + 4 \sqrt{2} v_{23}) + 2 \epsilon\Bigg]\\ & + 2 c_\theta s_\theta \Bigg[2 m_K^2 (\sqrt{2} - v_{23}) - 2 m_\pi^2 (\sqrt{2} - v_{23}) + \sqrt{2} \epsilon\Bigg]\Bigg\} + s_\theta (s_\theta + 2 c_\theta v_{23}) \Lambda_1,
\end{aligned}
\end{equation}
\begin{equation}
\begin{aligned}
\delta^{\eta'}_{k}=&\frac{8 L_5}{3 F^2} \Bigg\{c_\theta^2 \Bigg[m_K^2 (2 - 4 \sqrt{2} v_{23}) + m_\pi^2 (1 + 4 \sqrt{2} v_{23}) + \epsilon\Bigg]\\ & + s_\theta^2 \Bigg[4 m_K^2 (1 + \sqrt{2} v_{23}) - m_\pi^2 (1 + 4 \sqrt{2} v_{23}) + 2 \epsilon\Bigg]  \\ &-2 c_\theta s_\theta \Bigg[2 m_K^2 (\sqrt{2} - v_{23}) - 2 m_\pi^2 (\sqrt{2} - v_{23}) + \sqrt{2} \epsilon\Bigg]\Bigg\} + c_\theta (c_\theta - 2 s_\theta v_{23}) \Lambda_1,
\end{aligned}
\end{equation}
\begin{equation}
\begin{aligned}
\delta^{\eta\eta'}_{m^{2}}=&-\frac{32L_8}{3 F^2}
\Bigg\{c_\theta s_\theta \Bigg[-2 m_K^4 (1 + 4 \sqrt{2} v_{23}) + 2 m_K^2 (m_\pi^2 + 4 \sqrt{2} m_\pi^2 v_{23} - \epsilon) + 
m_\pi^2 \epsilon\Bigg]  \\ &+ s_\theta^2 \Bigg[-2 m_K^4 (\sqrt{2} - v_{23}) + \sqrt{2} m_\pi^2 \epsilon + 
2 m_K^2 \big[ m_\pi^2 (\sqrt{2} - v_{23}) - \sqrt{2} \epsilon\big]\Bigg]  \\ &+ c_\theta^2 \Bigg[2 m_K^4 (\sqrt{2} - v_{23}) - \sqrt{2} m_\pi^2 \epsilon + m_K^2 \big[-2 m_\pi^2 (\sqrt{2} - v_{23}) + 2 \sqrt{2} \epsilon\big]\Bigg]\Bigg\}  \\ &- \frac{1}{3} \Bigg\{2 c_\theta s_\theta \Bigg[m_K^2 (2 - 4 \sqrt{2} v_{23}) + m_\pi^2 (1 + 4 \sqrt{2} v_{23}) + \epsilon\Bigg] + c_\theta^2 \Bigg[-2 m_\pi^2 (\sqrt{2} - v_{23})  \\ &+ 2 m_K^2 (\sqrt{2} + 2 v_{23}) +\sqrt{2} \epsilon\Bigg] - s_\theta^2 \Bigg[-2 m_\pi^2 (\sqrt{2} - v_{23}) + 2 m_K^2 (\sqrt{2} + 2 v_{23}) +\sqrt{2} \epsilon\Bigg]\Bigg\} \Lambda_2,
\end{aligned}
\end{equation}
\begin{equation}
\begin{aligned}
\delta_{m_{\eta}^{2}}=&\frac{16 L_8}{3F^2}\Bigg\{c_\theta^2 \Bigg[3 m_\pi^4 + 8 m_K^4 (1 + \sqrt{2} v_{23}) - 8 m_K^2 (m_\pi^2 + \sqrt{2} m_\pi^2 v_{23} - \epsilon) - 
4 m_\pi^2 \epsilon\Bigg] \\ &+ s_\theta^2 \Bigg[3 m_\pi^4 + m_K^4 (4 - 8 \sqrt{2} v_{23}) - 2 m_\pi^2 \epsilon + 4 m_K^2 \big[m_\pi^2 (-1 + 2 \sqrt{2} v_{23}) + \epsilon\big]\Bigg]\\ & + 4 c_\theta s_\theta \Bigg[2 m_K^4 (\sqrt{2} - v_{23}) - \sqrt{2} m_\pi^2 \epsilon + m_K^2 \big[-2 m_\pi^2 (\sqrt{2} - v_{23}) + 2 \sqrt{2} \epsilon\big]\Bigg]\Bigg\} \\ &+ \frac{2}{3}\Bigg\{ 2 \sqrt{2}c_\theta^2 (m_K^2 - m_\pi^2) v_{23} + s_\theta^2 \Bigg[m_K^2 (2 - 2 \sqrt{2} v_{23}) \\ &+ m_\pi^2 (1 + 2 \sqrt{2} v_{23}) + \epsilon\Bigg]+ c_\theta s_\theta \Bigg[-2 m_\pi^2 (\sqrt{2} - v_{23}) + 2 m_K^2 (\sqrt{2} + 2 v_{23}) + \sqrt{2} \epsilon\Bigg] \Bigg\}\Lambda_2,
\end{aligned}
\end{equation}
\begin{equation}\label{eq.enddiff}
\begin{aligned}
\delta_{m_{\eta'}^{2}}=&\frac{16L_8}{3F^2}\Bigg\{s_\theta^2 \Bigg[3 m_\pi^4 + 8 m_K^4 (1 + \sqrt{2} v_{23}) - 8 m_K^2 (m_\pi^2 + \sqrt{2} m_\pi^2 v_{23} - \epsilon) - 4 m_\pi^2 \epsilon\Bigg] \\ &+ c_\theta^2 \Bigg[3 m_\pi^4 + m_K^4 (4 - 8 \sqrt{2} v_{23}) - 2 m_\pi^2 \epsilon + 4 m_K^2 \big[m_\pi^2 (-1 + 2 \sqrt{2} v_{23}) + \epsilon\big]\Bigg] \\ &+ 4 c_\theta s_\theta \Bigg[-2 m_K^4 (\sqrt{2} - v_{23}) + \sqrt{2} m_\pi^2 \epsilon + 2 m_K^2 \big[m_\pi^2 (\sqrt{2} - v_{23}) - \sqrt{2} \epsilon\big]\Bigg]\Bigg\}\\ & + \frac{2}{3}\Bigg\{ 2 \sqrt{2} (m_K^2 - m_\pi^2) s_\theta^2 v_{23} + c_\theta^2 \Bigg[m_K^2 (2 - 2 \sqrt{2} v_{23})\\ & + m_\pi^2 (1 + 2 \sqrt{2} v_{23}) + \epsilon\Bigg] - c_\theta s_\theta\Bigg[-2 m_\pi^2 (\sqrt{2} - v_{23}) + 2 m_K^2 (\sqrt{2} + 2 v_{23}) + \sqrt{2} \epsilon\Bigg]\Bigg\}\Lambda_2 ,
\end{aligned}
\end{equation}

\begin{equation}\label{eq.beginsame}
\begin{aligned}
\delta^{\pi\eta}_{k}=&\dfrac{-8L_{5}}{3F^{2}}\bigg\{2v_{12}\big[m^{2}_{K}(2 c^{2}_\theta+2\sqrt{2}c_\theta s_\theta+s^{2}_\theta)-m^{2}_{\pi}(2+2\sqrt{2}c_\theta s_\theta-s^{2}_\theta)\big]+\sqrt{3}\epsilon(\sqrt{2}s_\theta-c_\theta)\\ &-2v_{13}(m^{2}_{K}-m^{2}_{\pi})(\sqrt{2}c^{2}_\theta-c_\theta s_\theta-\sqrt{2}s^{2}_\theta)
\bigg\}+\Lambda_{1}s_\theta(c_\theta v_{13}-s_\theta v_{12}),
\end{aligned}
\end{equation}

\begin{equation}
\begin{aligned}
\delta^{\pi\eta}_{m^{2}}=&-\dfrac{16L_{8}}{3F^{2}}\big[4v_{12}m^{2}_{K}(m^{2}_{K}-m^{2}_{\pi})(2c^{2}_\theta+2\sqrt{2}c_\theta s_\theta+s^{2}_\theta)-4v_{13}m^{2}_{K}
(m^{2}_{K}-m^{2}_{\pi})(\sqrt{2}c^{2}_\theta-c_\theta s_\theta\\&-\sqrt{2}s^{2}_\theta)-2\sqrt{3}m^{2}_{\pi}\epsilon(c_\theta-\sqrt{2}s_\theta)\big]-\dfrac{\Lambda_{2}}{3}
\bigg\{2v_{12}s_\theta\big[2m^{2}_{K}(\sqrt{2}c_\theta+s_\theta)+m^{2}_{\pi}(s_\theta-2\sqrt{2}c_\theta)\big]\\&+2v_{13}\big[m^{2}_{K}(-\sqrt{2}c^{2}_\theta-2c_\theta s_\theta+\sqrt{2}s^{2}_\theta)+m^{2}_{\pi}(\sqrt{2}c^{2}_\theta-c_\theta s_\theta-\sqrt{2}s^{2}_\theta)\big]+\sqrt{6}s_\theta\epsilon\bigg\}\,,
\end{aligned}
\end{equation}

\begin{equation}
\begin{aligned}
\delta^{\pi{\eta}'}_{k}=&\dfrac{8L_{5}}{3F^{2}}\bigg\{2v_{12}(m^{2}_{K}-m^{2}_{\pi})(\sqrt{2}c^{2}_\theta-c_\theta s_\theta-\sqrt{2}s^{2}_\theta)+\sqrt{3}\epsilon(\sqrt{2}c_\theta+s_\theta)+2v_{13}\big[-m^{2}_{K}(c^{2}_\theta\\&-2\sqrt{2}c_\theta s_\theta +2s^{2}_\theta)
+m^{2}_{\pi}(2-2\sqrt{2}c_\theta s_\theta-c^{2}_\theta)\big]\bigg\}+\Lambda_{1}c_\theta(-c_\theta v_{13}+s_\theta v_{12})\,,
\end{aligned}
\end{equation}

\begin{equation}
\begin{aligned}
\delta^{\pi{\eta}'}_{m^{2}}=&\dfrac{16L_{8}}{3F^{2}}\big[4v_{12}m^{2}_{K}(m^{2}_{K}-m^{2}_{\pi})(\sqrt{2}c^{2}_\theta-c_\theta s_\theta-\sqrt{2}s^{2}_\theta)
-4v_{13}m^{2}_{K}(m^{2}_{K}-m^{2}_{\pi})(c^{2}_\theta-2\sqrt{2}c_\theta s_\theta+2s^{2}_\theta)\\ &+2\sqrt{3}m^{2}_{\pi}\epsilon(\sqrt{2}c_\theta+s_\theta)\big]
+\dfrac{\Lambda_{2}}{3}\bigg\{2v_{13}c_\theta\big[2m^{2}_{K}(-c_\theta+\sqrt{2}s_\theta)-m^{2}_{\pi}(c_\theta+2\sqrt{2}s_\theta)\big]
\\ &+2v_{12}\big[m^{2}_{K}(\sqrt{2}c^{2}_\theta+2c_\theta s_\theta-
\sqrt{2}s^{2}_\theta)+m^{2}_{\pi}(-\sqrt{2}c^{2}_\theta+c_\theta s_\theta+\sqrt{2}s^{2}_\theta)\big]+\sqrt{6}c_\theta\epsilon\bigg\}\,,
\end{aligned}
\end{equation}

\begin{equation}\label{eq.deltakpi}
\begin{aligned}
\delta^{\pi}_k=\dfrac{8L_{5}m^{2}_{\pi}}{F^{2}}\,,\quad \delta_{m^{2}_{\pi}}=\dfrac{16L_{8}m^{4}_{\pi}}{F^{2}}\,,
\quad \delta^{K}_k=\dfrac{8L_{5}m^{2}_{K}}{F^{2}}\,, \quad \delta_{m^{2}_{K}}=\dfrac{16L_{8}m^{4}_{K}}{F^{2}}\,.
\end{aligned}
\end{equation}

\bibliography{agg-IB}
\bibliographystyle{apsrev4-2}

\end{document}